\documentclass{aastex61}
\usepackage{graphicx}
\usepackage{placeins}
\usepackage{amsmath}
\newcounter{subfigure}

\begin{document}

\title{ Exploring the Dust Content of Galactic Winds with {\em
    Herschel}. \\II. Nearby Dwarf Galaxies\footnote{{\em Herschel} is
    an ESA space observatory with science instruments provided by
    European-led Principal Investigator consortia and with important
    participation from NASA.}}
	
\author{Alexander McCormick} \affiliation{Department of Physics,
  University of South Florida, Tampa, FL 33620, USA}

\author{Sylvain Veilleux} \affiliation{Department of Astronomy and
  Joint Space-Science Institute, University of Maryland, College Park,
  MD 20742}

\author{Marcio Mel\'endez} \affiliation{NASA Goddard Space Flight
  Center, Greenbelt, MD 20771, USA ; Wyle Science, Technology and
  Engineering Group, 1290 Hercules Avenue, Houston, TX 77058, USA}

\author{Crystal L. Martin} \affiliation{Department of Physics,
  University of California, Santa Barbara, CA 93106, USA}

\author{Joss Bland-Hawthorn} \affiliation{Department of Physics,
  University of Sydney, Sydney, NSW 2006, Australia}

\author{Gerald Cecil} \affiliation{Department of Physics, University
  of North Carolina, Chapel Hill, NC 27599, USA}

\author{Fabian Heitsch} \affiliation{Department of Physics, University
  of North Carolina, Chapel Hill, NC 27599, USA}

\author{Thomas M{\"u}ller} \affiliation{Max-Planck-Institute for
  Extraterrestrial Physics (MPE), Giessenbachstrasse 1, 85748
  Garching, Germany}

\author{David S. N. Rupke} \affiliation{Department of Physics, Rhodes
  College, Memphis, TN 38112, USA}

\author{Chad Engelbracht} \affiliation{Department of Astronomy,
  University of Arizona, Tucson, AZ 85721, USA (deceased)}

\begin{abstract}
We present the results from an analysis of deep {\em Herschel Space Observatory} observations of six nearby dwarf galaxies known to host galactic-scale winds. The superior far-infrared sensitivity and angular resolution of {\em Herschel} have allowed detection of cold circumgalactic dust features beyond the stellar components of the host galaxies traced by {\em Spitzer} 4.5 $\mu$m images. Comparisons of these cold dust features with ancillary data reveal an imperfect spatial correlation with the ionized gas and warm dust wind components. We find that typically $\sim$10-20\% of the total dust mass in these galaxies resides outside of their stellar disks, but this fraction reaches $\sim$60\% in the case of NGC~1569. This galaxy also has the largest metallicity (O/H) deficit in our sample for its stellar mass. Overall, the small number of objects in our sample precludes drawing strong conclusions on the origin of the circumgalactic dust. We detect no statistically significant trends with star formation properties of the host galaxies, as might be expected if the dust were lifted above the disk by energy inputs from on-going star formation activity.  Although a case for dust entrained in a galactic wind is seen in NGC~1569, in all cases, we cannot rule out the possibility that some of the circumgalactic dust might be associated instead with gas accreted or removed from the disk by recent galaxy interaction events, or that it is part of the outer gas-rich portion of the disk that lies below the sensitivity limit of the {\em Spitzer} 4.5 $\mu$m data.
\end{abstract}



\section{INTRODUCTION}
\label{intro}

Galactic winds driven by stellar processes and active galactic nuclei
(AGN) are a fundamental mechanism of galaxy evolution (see
\citealt{vei05} for a review). Simulations have shown that the
outflows of material in winds can inhibit the growth of central
supermassive black holes and curb the star formation rates (SFR) in
galaxies \citep[e.g.,][and references therein]{naa17}.  Winds have also been
invoked to explain a host of galaxy observations, including the
mass-metallicity relation \citep[e.g.,][]{tre04,man10,lar10,kas16,tel16,bar17,san17} and the
relation between the central black hole mass and the bulge velocity
dispersion \citep[e.g.,][]{fer00,geb00,gul09,kor13}.
Furthermore, the ubiquity of galactic winds at $z$ \textgreater \ 1
\citep[e.g.,][and references therein]{wei09,ste10,sha11,erb12,mar12}
points to their importance in understanding the past history of the
universe. Therefore, detailed observations of galactic winds are
critical to fleshing out the narrative of galaxy evolution. Although
negative feedback almost certainly assert greater influence at
high redshift, where strong starbursts and active galactic nuclei are
more commonplace, nearby sources provide the best opportunities for
detailed observations of the galactic winds. Dwarf galaxies are
particularly important in this context since their evolution is highly
sensitive to the effects of galactic winds and feedback in general 
\citep[e.g.,][]{vog14,mur15,sch15,dav17}.

Prior to 2005, much of the observational data on galactic winds
emphasized the entrained ionized material
\citep[e.g.,][]{hec90,leh95,vei05}, the neutral gas
\citep[e.g.,][]{hec00,rup02,rup05a,rup05b,rup05c,sch04,mar05}, and the
highly ionized X-ray emitting plasma 
\citep[e.g.,][]{rea97,mar99,pie00,mar02,mcd03,ehl04,huo04,str04a,str04b}. 
Only a few observations showed evidence that these outflows also
entrained dust \citep[e.g.,][]{alt99,hec00,rad01,tac05} and molecular
gas \citep[e.g.,][]{cur98,wal02}. This state of affairs has changed
dramatically in recent years, largely driven by the new capabilities
of space-borne infrared observatories
\citep[e.g.,][]{eng06,kan09,fis10,kan10,rou10,stu11,mcc13,vei13,mel15} and ground-based
near-infared imagers and spectrometers and mm-wave interferometers
\citep[e.g.,][]{sak06,vei09,fer10,irw11,ala11,aal12,bol13,rup13b,cic14,ala15,fer15,lin16,vei17}.
These new results have in turn triggered a flurry of theoretical
studies and numerical simulations of the interaction of winds with the
denser ISM aimed at understanding the nature and impact of these dusty
neutral and molecular outflows on galaxy evolution
\citep[e.g.,][]{coo08,coo09,fau12,zub13,zub14,mcc15,nim15,sca15,fer16,bru16,ban16,tan16,tan17,ric17,ric18}.

Dust grains found in the interstellar medium (ISM) were formed in the
atmospheres of evolved stars or during the outbursts of novae and
supernovae, but can also be destroyed and reconstituted in the
ISM. Interstellar dust plays a crucial role in galaxy evolution and
star formation. It absorbs and scatters light, it can act as the
catalyst to form molecules through reactions not possible in the gas
phase, it can enrich or deplete the interstellar gas via destruction
and evaporation of grains or condensation on grains, and electrons
liberated from dust grains via the photoelectric effect can heat the
ISM gas \citep{kru02,leq05}. Therefore, investigating the
distribution, mass, and energy of the dust in wind galaxies provides
critical information for understanding galaxy evolution. Observations
of dusty winds may even catch galaxies in the act of expelling their
star formation fuel, eventually halting stellar nurseries.

In the first paper of this series \citep[hereafter Paper I]{mel15}, we
presented {\em Herschel Space Observatory} \citep{pil10} observations
of the nearby wind galaxy NGC~4631. In this second paper of the
series, we examine six nearby dwarf galaxies selected to host known
galactic winds. The new results are based on an analysis of our deep
observations of these objects with {\em Herschel}. The angular
resolution and sensitivity of these data, far superior to previous
far-infrared (FIR) data obtained with {\em Spitzer} MIPS, allow us to
detect and characterize cool dust that lies outside the extent of the
stellar component of the host galaxy (hereafter called
``circumgalactic dust'').

In \S~\ref{sample}, we describe our sample, the methods of acquisition
of the {\em Herschel} data, and the ancillary data. In \S~\ref{reduce}
and~\ref{analysis}, we explain the process used to reduce the {\em
  Herschel} data and our analysis of these data, respectively. We
present our results of this analysis in \S~\ref{results_discuss},
including maps of each galaxy in {\em Herschel's} 70, 160, 250, 350,
and 500 $\mu$m channels, and comparisons of the basic properties of
the circumgalactic dust with those of the host galaxies. Our results
are summarized in \S~\ref{summary}. Appendix~\ref{appendix} contains a
short discussion of each galaxy.


\section{SAMPLE AND DESCRIPTION OF THE DATA}
\label{sample}

The six nearby dwarf galaxies in our sample were selected based on
their proximity (less than $\sim$10 Mpc), their stellar masses (less
than $\sim$10$^{10}$ M$_\odot$), and on previous evidence of galactic
winds at other wavelengths. Table~\ref{tbl:sample} lists the galaxies,
some of their properties, and selected references to evidence of
galactic winds for each source. This sample is not homogeneous -- it
spans a broad range of properties. All of these galaxies are actively
forming stars but one of them, He~2-10, likely also hosts an AGN
\citep{rei11,rei16}. He~2-10 is represented with an open circle in
some of the figures to differentiate it from the galaxies hosting
purely star formation. We also note that the membership of NGC~3077 in
the M81 group complicates the interpretation of the circumgalactic
material, so this object is also represented with a different symbol
in the figures and is treated separately in the discussion of the
results.

We acquired very deep (\textgreater \ 6 hrs) {\em Herschel Space
Observatory} PACS \citep{pog10} images in the 70 and 160 $\mu$m
channels for each of the six objects in our sample as part of a
two-cycle observational program to investigate the circumgalactic dust
of nearby star-forming galaxies in general. The long exposure times for
our very deep PACS images were necessary to detect the faint circumgalactic
dust emission and distinguish it from point spread function artifacts near the
peak of the infrared spectral energy distribution. Each of the 70 and 160
$\mu$m PACS observations consisted of a four-times repeated scan map
of either 30 or 40 scan legs 3$^\prime$ in length and separated by
4$^{\prime\prime}$ to fully map the galaxies and to obtain a very high
and homogeneous coverage well beyond the stellar extent of the
galaxies. These two-band scan-map observations were conducted seven
times, each time at a different angle with respect to the detector
array orientation. This technique virtually eliminates systematic
noise from low-level striping and reaches approximately Poisson noise
limits. We also acquired SPIRE \citep{gri10} data at 250, 350, and 500
$\mu$m for the one galaxy in our sample which had not already been
observed (NGC~1800) and downloaded the archived SPIRE data for the
other five galaxies. The dominant source of noise in deep SPIRE data
is confusion noise, i.e. confusion with unresolved foreground and background
sources, which does not depend on exposure time. SPIRE reaches this
confusion limit in exposure times of $\sim$0.6 hr or less (it depends on the exact
wavelength); hence the exposure times of the SPIRE data never exceed
this limit. The SPIRE observations were performed in the LargeScanMap
mode using orthogonal scan directions and multiple iterations for four out
of the six galaxies (single iterations for He~2-10 and NGC~1569). The
{\em Herschel} data are summarized in Table~\ref{tbl:data}. The {\em Herschel}
observations were performed between January 2012 and March 2013.

In addition to the new {\em Herschel} data, we also brought together
ancillary data for each galaxy in four additional bands: H$\alpha$,
4.5 $\mu$m, 8.0 $\mu$m, and 24 $\mu$m. The ancillary data are
summarized in Table~\ref{tbl:ancdata} and discussed in more detail
below.


\section{DATA REDUCTION}
\label{reduce}

Once the {\em Herschel} observations were completed, we extracted the
new PACS and SPIRE data from the {\em Herschel} Science Archive
(HSA). We downloaded the archival {\em Spitzer} Infrared Array Camera
(IRAC; 4.5 and 8.0 $\mu$m) and Multiband Imaging Photometer (MIPS; 24
$\mu$m) data from the NASA/IPAC Infrared Science Archive. Before
performing any reduction, we briefly examined the new {\em Herschel}
data with version 10.3.0 of the Herschel Interactive Processing
Environment (HIPE; \citealt{ott10}), and we also inspected the
H$\alpha$ data and pipeline-processed {\em Spitzer} data (PBCDs) using
SAOImage DS9.

\subsection{{\em Herschel} Data Reduction}
\label{H_reduce}

Following the steps in \cite{mel15}, we reprocessed our {\em
  Herschel} data with HIPE up to Level-1 employing the PACS photometer
pipeline \citep{bal14} and then passed these data on to the {\em
  Scanamorphos} v21 software \citep{rou13}. Since our deep {\em
  Herschel} observations aimed to detect very faint emission from cold
dust in circumgalactic regions, we employed {\em Scanamorphos}, which
was built specifically to handle scan mode observations like ours, the
preferred acquisition mode for nearby galaxies. {\em Scanamorphos}
exploits the redundancy of the observations in order to subtract low
frequency noise due to thermal and non-thermal components. It also
masks high frequency artifacts like cosmic ray hits before projecting
the data onto a map. The final pixel sizes, which are $\sim$ 1/4 the
size of the point-spread functions (PSFs) for each detector, are 1.4,
2.85, 4.5, 6.25, and 9.0$^{\prime\prime}$ for the 70, 160, 250, 350,
and 500 $\mu$m maps, respectively. The {\em Herschel} maps are shown
in Figures~\ref{fig:He2-10} -~\ref{fig:n5253} along with the ancillary
H$\alpha$, 4.5, 8.0, and 24 $\mu$m data.

\subsection{{\em Spitzer} Data Reduction}

Our reduction of the {\em Spitzer} IRAC data followed the same
procedures as those described in \cite{mcc13}, summarized
here. Starting from the basic calibrated data, we corrected any
electronic and optical banding in the 8.0 $\mu$m channel. In several of
our sample galaxies, the bright nuclei generate broad point spread
functions (PSFs) in both the 4.5 and 8.0 $\mu$m channels, which
overlap with circumgalactic regions, so we subtracted the wings of any
broad PSFs using the APEX and APEX QA modules of the {\em Spitzer
  Science Center}-provided MOPEX software \citep{mak05}. The 4.5
$\mu$m IRAC channel provides a better representation of the stellar
emission than the 3.6 $\mu$m IRAC channel, since polycyclic
aromatic hydrocarbon (PAH) emission from star formation in the galaxy
disks may contaminate the 3.6 $\mu$m IRAC channel
-- see Figure 1 in \cite{rea06}. Where
necessary, we also performed a background subtraction to better
distinguish background flux from circumgalactic features. Once the
basic calibrated data were processed, we created maps of both channels
using MOPEX. For NGC~1569, NGC~1705, and NGC~5253, we used the
previously-reduced data presented in \cite{mcc13}.


\section{DATA ANALYSIS}
\label{analysis}

\subsection{Disk-Circumgalactic Decomposition}
\label{disk}

Differentiating between a galaxy and its circumgalactic or extraplanar
or halo region necessitates defining an edge or border between the two
components when analyzing an image projected on the sky. This process
is necessarily subjective since the transition between the host galaxy
and its halo is smooth and gradual. It is particularly important to
keep this in mind when dealing with dwarf galaxies since they often
have H {\sc i} disks or structures that extend beyond their stellar
components \citep[e.g.,][]{ash14}. Here, we defined an edge by
choosing an isophote of 0.1 MJy~sr$^{-1}$ in the predominantly stellar
emission of the IRAC 4.5 $\mu$m channel map of each galaxy in our
sample.  This surface brightness threshold was selected to be faint
enough to represent the outer boundary of the stellar emission, but
sufficiently bright so that this boundary is not sensitive to the noise
structure (this surface brightness threshold is greater than 4$\sigma$
above the background noise for each map).  This limit effectively
corresponds to a uniform stellar mass surface density threshold across
our sample. Using this method does not account for the potential
occultation or coincidence of circumgalactic and disk emission due to
inclination. However, any inclination effect will tend to reduce the
amount of circumgalactic emission, not increase it. Therefore we can
consider these regions as setting an effective lower limit on the
circumgalactic emission. For the sake of consistency and simplicity,
we will call the regions encompassed within this 4.5 $\mu$m surface
brightness threshold ``disk'' regions, although most of the galaxies
in our sample do not exhibit a classical disk morphology. The disk and
circumgalactic regions defined in this fashion are shown in
Figures~\ref{fig:He2-10_CLEAN} $-$~\ref{fig:n5253_CLEAN}. The
circumgalactic flux region for NGC~1569 is necessarily conservative
because of the proximity of Galactic cirrus emission which are noticeably
impinging on the circumgalactic region in the SPIRE maps (see
Figures~\ref{fig:n1569} and \ref{fig:n1569_CLEAN}).

\subsection{Removal of the Disk PSFs with a CLEAN Algorithm}

In both the PACS and SPIRE instruments, the PSFs are broad enough that
the wings from bright disk sources bleed out and contribute excess
flux to the circumgalactic regions where we are looking for just the
faint circumgalactic dust emission. To account for this excess, we
employed a modified version of the CLEAN algorithm\footnote{Adapted
  from http://www.mrao.cam.ac.uk/$\sim$bn204/alma/python-clean.html.}
\citep{hog74}. Our CLEAN algorithm finds the peak pixel within an area
similar to the disk region, subtracts the appropriate PSF scaled by a
pre-defined gain as a fraction of the peak pixel value, and repeats
these two steps until the peak pixel value meets or drops below a
pre-defined minimum threshold value. Once the minimum threshold value
is reached, the algorithm outputs component and residual images. Since
the galaxies in our sample contain broad, bright areas within their
disk regions, we chose observations of Vesta (Obs.\ IDs 1342195624,
1342195625, also the reference PSF for PACS) and Uranus (Obs.\ ID
1342197342) as our beam PSFs. We selected {\em Herschel} observations
of Vesta and Uranus with the same scan speed as our observations,
which is important for matching the shape of the PSFs. We processed
the Vesta and Uranus PSFs through the same combination of HIPE and
{\em Scanamorphos} as described in \S~\ref{H_reduce}. Next, we rotated
the PSFs to match the galaxy observations, centered the PSFs on their
central pixels, and scaled the PSFs by normalizing their central peak
pixels to a value of 1 for gain multiplication. Finally, we applied
our CLEAN algorithm to each of the PACS and SPIRE images iteratively
lowering the threshold value to determine circumgalactic flux value
convergence. Figure~\ref{fig:CLEAN} illustrates the ``before'' and
``after'' results of applying our CLEAN algorithm to the maps of
NGC~1569. Across all bands before application of the CLEAN algorithm,
we find that $\sim$75\% (standard deviation $\approx$ 20\%) of the
flux in the circumgalactic region has bled out from the disk
region. There is no apparent trend with wavelength. Residual maps
after application of our CLEAN algorithm are shown in
Figures~\ref{fig:He2-10_CLEAN} $-$~\ref{fig:n5253_CLEAN}.

\subsection{{\em Herschel} Flux Measurements}
\label{flux_measure}

In each of the five {\em Herschel} bands, we took global flux
measurements from the original {\em Scanamorphos} maps using a
circular or elliptical aperture centered on the galaxy and containing
the most extended circumgalactic cold dust features. Note that the
most extended features did not always appear in the same {\em
  Herschel} band. In each of the five bands, we calculated the cleaned
circumgalactic flux by adding up the flux in the CLEAN residual maps
outside of the disk region out to the boundary of the region used for
the global flux measurements. We then calculated the corrected disk
flux by subtracting the cleaned circumgalactic flux from the original
global flux value. Using this method, the excess flux that bled out of
the disk region into the circumgalactic region is added back into the
disk flux total.

We also took sample flux measurements in darker, mostly source-free
sky regions outside the global flux regions to confirm whether {\em
  Scanamorphos} had properly removed the background in the {\em
  Herschel} maps. The background regions in the PACS maps contained
flux values close enough to zero as to be indistinguishable from
noise. However, the backgrounds of the SPIRE maps still contained
enough flux from either background sources or residual noise to
require a background correction. We measured the flux in background
regions identical in size and shape to the global region tiled around
the global region perimeter. The bottom row of panels in each of
Figures~\ref{fig:He2-10_CLEAN} -~\ref{fig:n5253_CLEAN} shows the SPIRE
background regions. We then averaged these background fluxes and
subtracted the average to get the final SPIRE global flux
measurement. The SPIRE background contribution fell somewhere in the
2-14\% range (mean $\sim$ 8\%) with NGC~1569 as an outlier at about
35\% (weighted average of the regions shown in
Figure~\ref{fig:n1569_CLEAN}), likely due to some contribution from
nearby Galactic cirrus emission.

We adopted a conservative flux calibration uncertainty of 10\%, which
consists of the systematic (4-5\%), statistical (1-2\%), and PSF/beam
size (4\%) uncertainties. The aperture sizes of the global regions in
our sample had radii $\sim$ 110$^{\prime\prime}$. Derived aperture
corrections of order $\sim$ 3-4\% \cite[Paper I,][]{bal14} are small
compared to this calibration uncertainty. Color
corrections\footnote{http://herschel.esac.esa.int/twiki/pub/Public/PacsCalibrationWeb/cc\_report\_v1.pdf}
are on the order of 2-3\% (for sources with a temperature of $\sim$ 30
K), which are also small compared with our 10\% flux uncertainty. For
sources with temperatures closer to 20 K, the color correction to the
70 $\mu$m PACS flux is 8\%. The global, disk, and circumgalactic flux
values are listed in Table~\ref{tbl:flux}.

\subsection{Spectral Energy Distribution (SED) Fitting}
\label{sed_fit}

In order to characterize the cold dust of each galaxy in our sample,
we fit a modified blackbody (MBB) following the one found in Section
3.1 of \cite{smi12}. We fit the sets of 70, 160, 250, 350, and 500
$\mu$m global and disk flux values to:

\begin{equation}
\label{eq:sed_fit}
S_{\nu} \ = \ \frac{M_d \ \kappa_{\nu} \ B_{\nu} (T_d)}{d^2},
\end{equation}

\noindent where $M_d$ and $T_d$ are the dust mass and temperature,
respectively, $d$ is the distance to the galaxy listed in
Table~\ref{tbl:sample}, $B_{\nu}$ is the Planck function, and
$\kappa_{\nu}$ is the dust absorption coefficient which has a power
law dependence with dust emissivity index $\beta$ where $\kappa_{\nu}$
= $\kappa_0$($\nu$/$\nu_0$)$^{\beta}$. The constant $\kappa_0$ is the
dust opacity at $\nu_0$ = 350 $\mu$m: 0.192 m$^2$ kg$^{-1}$
\citep{dra03}. We fit the parameters $M_d$, $T_d$, and $\beta$ to each
set of global and disk fluxes to derive $M_{global}$, $T_{global}$,
$M_{disk}$, $T_{disk}$. The fits also included the 10\% flux
calibration uncertainties. Fitting all three parameters often produced
unphysical $\beta$ values, so instead we fit just $M_d$ and $T_d$ and
adopted a fixed value of $\beta$ = 2.0, which is indexed to the dust
opacity $\kappa_0$. Since the 70 $\mu$m flux value likely includes a
contribution from a warmer dust component \citep{cas12}, we treated it
as an upper limit in our MBB fits. For consistency, we did not include
the 100 $\mu$m flux values in the MBB fits, since these observations have
not been done to match the depth of our 70 and 160 $\mu$m data. We
estimated the uncertainties in the fit parameters by running 500 Monte
Carlo MBB fits for each set of fluxes, which we let vary randomly
within the associated uncertainties for each
run. Figures~\ref{fig:mbb_global} and~\ref{fig:mbb_disk} show the
global and disk MBB fits, respectively, with the disk MBB fits shown
for fluxes both before and after application of the CLEAN
algorithm. The MBB fit parameters and the uncertainties estimated with
the Monte Carlo method are listed in Table~\ref{tbl:sed}.

We also tried fitting the circumgalactic flux values to the same
single MBB, but these fits produced unphysical results (e.g.,
unrealistically high dust masses). We believe that this is due to the
faint flux of the circumgalactic emission and the heterogeneous nature
of this emission (e.g., filaments of different temperatures and masses
which are not well represented by a single MBB). Perhaps a detailed
superposition of MBBs could reproduce the combined circumgalactic SED,
but that avenue of analysis wades into more degenerate, subjective
territory. To avoid this, we calculated the circumgalactic cold dust
masses by simply using $M_{cg}$ = $M_{global}$ $-$ $M_{disk}$.

The properties of the cold dust in the disk and circumgalactic regions
depend on the value of the 4.5 $\mu$m surface brightness we adopted as
the boundary between the disk and circumgalactic regions. For
instance, choosing a higher 4.5 $\mu$ surface brightness of 0.3 MJy
sr$^{-1}$ instead of the favored value of 0.1 MJy sr$^{-1}$ for the
disk -- CGM boundary produces a decrease of $\sim$ 10\% in the disk
masses as well as an increase of $\sim$ 1 K in the disk temperatures
when fitting the MBBs. However, since the disk masses ($M_{disk}$) are
similar in magnitude to the global masses ($M_{global}$) and we used
$M_{cg}$ = $M_{global}$ - $M_{disk}$, a small decrease in the disk
mass produces a proportionally larger increase in the circumgalactic
mass ($M_{cg}$). With a surface brightness threshold of 0.3 MJy
sr$^{-1}$, we found circumgalactic masses $\sim$ 50-100\% more massive
than with the more conservative regions we have chosen. Therefore our
choice of the larger, more conservative disk region does not
significantly affect the derived disk properties, but does produce a
more conservative circumgalactic dust mass.


\section{RESULTS AND DISCUSSION}
\label{results_discuss}

\subsection{Morphology}

We find circumgalactic dust features in all six galaxies of our
sample (see Figures~\ref{fig:He2-10_CLEAN} -~\ref{fig:n5253_CLEAN} and
Appendix~\ref{appendix}). The most extended features range from 1.2
kpc (NGC~1800) to 2.6 kpc (He~2-10), as measured from the center of
the galaxy or from the mid-plane of the disk-like region out to the
furthest part of the feature at the 3$\sigma$ level above the
background. These features extend to scales of $\sim$1.2-2.5 times the
radii of the stellar component as measured by the 4.5 $\mu$m stellar
disk region, and typically do not trace the orientation of the stellar
disk. The circumgalactic features vary in morphology including
extended filaments (NGC~1569, NGC~1705, NGC~1800, NGC~3077, and
NGC~5253), clouds or knots of dust apparently separated from the disk
region (NGC~1569, NGC~1705, and NGC~5253), as well as broader regions
extending over a large range of galactocentric angles (e.g.,
He~2-10). As a member of the M81 group, NGC~3077 is in the process of
interacting with M81 and M82, so much of the circumgalactic cold dust
features might be attributed to tidal stripping. In particular, the
features to the east, west and north-northeast of NGC~3077 are likely
due to tidal stripping since 21-cm observations
\citep{cot76,van79,yun94} show H {\sc i} tidal streams in those
regions.

Some circumgalactic cold dust features coincide with similar emission
features in the ancillary H$\alpha$, 8.0, and 24 $\mu$m data (e.g.,
filaments of NGC~1569 and NGC~1800), but others do not (e.g., the
broader plume extending SSW from NGC~1705). Figure~\ref{fig:halpha}
shows the 160 $\mu$m PACS maps overlaid with H$\alpha$ contours based
on the data listed in Table~\ref{tbl:ancdata}. The morphology of
the circumgalactic cold dust features does not always match the
morphology of their warmer counterparts at shorter wavelengths. For
example, the cold dust around NGC~1705 exhibits a somewhat filamentary
morphology with a broad plume extending SSW and spatially extended
emission to the north, while the H$\alpha$ features are more shell-like
as shown by the arcs of emission directly south and north-west of
the disk. The hot and warm wind components traced by H$\alpha$ and
PAH emission do not correlate exactly with the cold dust
component, and therefore should not be used as a predictor of cold
dust and vice versa. This result supports the idea of shielded regions
within a galactic wind where dust may remain protected from sputtering
via thermal, radiation, or collisional processes
\citep[e.g.,][]{jur98,gne14}. Appendix~\ref{appendix} contains further
discussion of individual galaxies, their cold dust features, and
comparisons with H {\sc i} and X-ray observations.

\subsection{70 $\mu$m / 160 $\mu$m Ratio Maps}

In order to estimate the dust temperature and its spatial distribution
using a proxy measurement, we made ratio maps where the 70 $\mu$m flux
maps were divided by the 160 $\mu$m flux maps.\footnote{The dust
  emissivity, $\beta$, and dust grain size distribution can also
  contribute to variations in 70 $\mu$m / 160 $\mu$m ratios (see Paper
  I).} We first convolved the 70 $\mu$m maps to the pixel scale of the
160 $\mu$m maps using the convolution kernels described in
\cite{ani11}. We then aligned the resulting convolved images with the
160 $\mu$m maps and took the ratio, limiting the plotted ratios to
pixels where the 160 $\mu$m map had values \textgreater 0.0001 Jy
pixel$^{-1}$ ($\sim$1$\sigma$ above the background) for the sake of
clearly displaying galactic features. The 70 $\mu$m / 160 $\mu$m ratio
maps are shown in Figure~\ref{fig:ratios}. All of the galaxies exhibit
warmer dust temperatures (ratio \textgreater 1.0) near their nuclei as
would be expected for the stronger radiation field in nuclear regions,
while their outer regions tend to appear cooler. Unlike the other
galaxies, NGC~3077 has a striking temperature gradient along the
northeast to southwest axis. Observations have shown that the tidal
features of NGC~3077 are significantly cooler than the galaxy itself
\citep{wal11}. The cooler dust to the southwest of the nucleus appears
to coincide with warm dust in the 8.0 and 24 $\mu$m maps and the H
{\sc i} tidal stream towards M81 \citep{cot76,van79,yun94}, but no
obvious H$\alpha$ feature. In general, the warmer regions in the ratio
maps shown by yellow and red colors in Figure~\ref{fig:ratios} do not
appear to correlate well in a qualitative sense with much of the warm
dust features in the 8.0 and 24 $\mu$m maps or hot ionized gas of the
H$\alpha$ maps, though some exceptions arise like the bright filament
to the southwest of NGC~1569 and the warmer region to the east of the
nucleus in NGC~5253 coinciding with the ionization cone reported by
\cite{zas11}. Finally, the temperature structure of He~2-10 is highly
peaked at the center, perhaps due to AGN activity in its nucleus
\citep{rei11,rei16}.

\subsection{Global Gas-to-Dust Ratios}

Table~\ref{tbl:sed} lists the cold dust masses and temperatures
derived by fitting MBB to the FIR emission from the global and disk
regions (\S~\ref{sed_fit}). The fitted global cold dust masses are in
the range $\sim$10$^{5}$ $-$ 10$^{6.5}$ $M_{\odot}$.  For comparison,
the total gas masses (H {\sc i} + H$_2$) for the galaxies in our
sample lie in the range $\sim$1.5 $-$ 9.0 $\times$ 10$^{8}$
$M_{\odot}$ (Table~\ref{tbl:sample}), implying gas-to-dust ratios of
$\sim$ 200 $-$ 1500. These elevated values compared with that of the
Milky Way (gas-to-dust ratio of $\sim$140; e.g., \citealt{dra07})
confirm recent results by \cite{rem14}, \cite{rem15}, and \cite{fel15}
that low-metallicity dwarf galaxies are deficient in dust compared to
our own Galaxy.

\subsection{Comparison with the Host Galaxy Properties}

As discussed in \S~\ref{sed_fit}, we calculated the circumgalactic
cold dust mass ($M_{cg}$) listed in Table~\ref{tbl:sed} by subtracting
the disk cold dust mass ($M_{disk}$) from the global cold dust mass
($M_{global}$). Figure~\ref{fig:cg_tot} shows $M_{cg}$ versus the
total baryonic (stellar plus gas) mass for the galaxies in our
sample. The stellar masses ($M_*$) are listed in
Table~\ref{tbl:sample} and were calculated using the same method as
\cite{zas13} (see Table~\ref{tbl:sample} for more detail). The
circumgalactic dust exhibits a weak (Pearson's $r$ = 0.80; P[null] =
0.055) positive linear correlation with the baryonic mass. This
correlation should be considered with caution since it is based on a
sample of only six objects (this issue is discussed in more detail
below).  If real, this positive correlation may reflect one of two
things: (1) that bigger, more massive galaxies simply have more of
everything, including stars, gas, and dust, inside and outside of
their stellar disks, or (2) there is actually a physical connection
between the deposition of the dust in the circumgalactic environment
via star formation-driven winds or tidal processes and the processes
associated with the stellar build-up of galaxies.

In an attempt to differentiate between these two scenarios, we plot
$M_{cg}$ normalized by $M_{global}$ versus the star formation rate
surface density ($\Sigma_{SFR}$, see Table~\ref{tbl:sample}) in
Figure~\ref{fig:cgmass_Sigma}. This last quantity is estimated using
the (infrared + H$\alpha$) star formation rate (based on the
prescription of \cite{ken09}) together with the extent of the
H$\alpha$ emission ($R_{H\alpha}$) listed in Table 1 of
\cite{cal10}.\footnote{Note again that an AGN may contribute to the
  far-infrared emission in He~2-10, so its star formation rate and
  global and circumgalactic dust masses carry that caveat, indicated
  by the open red circle in
  Figures~\ref{fig:cg_tot}-\ref{fig:cgmass_delZ}.}  Keeping in mind
that the typical dust mass uncertainties from the MBB fitting are on
the order of $\sim$ 40\%, Figure~\ref{fig:cgmass_Sigma} shows that
typically $\sim$10-20\% of the cold dust in these wind-hosting dwarf
galaxies lies in the circumgalactic region. NGC~1569 is an outlier,
having as much as $\sim$60\% of its cold dust in the circumgalactic
region. Overall, Figure~\ref{fig:cgmass_Sigma} shows no obvious trend,
contrary to what might be naively expected if the dust is lifted above
the disk by energy inputs from star formation activity in the disk. We
caution against drawing general conclusions about dwarf galaxies based
on a sample of only six objects. Furthermore, we note that the lack of
a trend in Figure~\ref{fig:cgmass_Sigma} is not unexpected if the
stellar mass $-$ metallicity relation in low-mass dwarf galaxies is
due to inefficient conversion of gas into stars in these objects
rather than supernova-driven outflows \citep[e.g.][]{tas08}.

Other factors may also be affecting our results.  One possibility is
that our measurements might be missing the very cold dust in giant
molecular clouds within the disk, thus over-estimating the fraction of
dust found in the circumgalactic region. However, this cannot explain
the large circumgalactic dust mass fraction measured in NGC~1569 since
the H$_2$ gas mass is only about 1\% of the H {\sc i} gas mass in this
object (see Table~\ref{tbl:sample}), and thus there are very few
hiding places for additional dust in the disk. A more valid concern
about NGC~1569 is contamination from Galactic cirrus emission. While
the sky immediately surrounding NGC~1569 appears mostly free of
Galactic cirrus in the SPIRE bands, there is indeed significant cirrus
emission in adjacent regions. Our estimate of the background and foreground flux attempts to account
for this potential contamination by including a region (upper right
ellipse in Figure~\ref{fig:n1569_CLEAN}) which includes some cirrus
emission. If foreground emission from Galactic cirrus happened to be
projected against the circumgalactic region of NGC~1569 and not the
disk region, then $M_{cg}$ / $M_{global}$ would be
overestimated. However, we find that this convoluted scenario is
unlikely to fully explain the large value of $M_{cg}$ / $M_{global}$
given the clear detection in Figure 3 of the circumgalactic dust
emission not only at 250, 350, and 500 $\mu$m but also at 70 and 160
$\mu$m, where cirrus emission is no longer an issue.

Another potential source of scatter in Figure~\ref{fig:cgmass_Sigma}
is the presence of dusty tidal streams contributing to the
circumgalactic far-infrared emission.  NGC~3077 is the clearest case
of a tidally disrupted dwarf in our sample, where this effect is likely
important. \cite{kar99} developed a tidal index parameter ($\Theta_i$)
as a quantitative measure of the importance of tidal effects based on
the masses and three-dimensional distances from its nearest
neighbors. The maximum value of $\Theta_i$ determines the galaxy's
``Main Disturber" (MD). A maximum $\Theta_i$ greater than zero
indicates gravitational interaction with the MD within a tidal radius,
while a value less than zero indicates a relatively isolated galaxy
outside that tidal radius. Amongst our sample, NGC~1569, NGC~3077, and
NGC~5253 have positive $\Theta_i$ values \citep{kar14}, of which
NGC~3077 unsurprisingly has the highest value. In contrast, He~2-10,
NGC~1705, and NGC~1800 all have negative $\Theta_i$, indicating their
isolation and thus likely negligible contributions from tidal streams
to the measured circumgalactic dust. Our data do not show any
systematic difference in $M_{cg}$ / $M_{global}$ between these two
groups of objects.

If the dust in a galaxy carries fractionally more metals than its gas,
then dust infall onto a galaxy or outflow from the galaxy can
significantly affect the overall ISM metallicity of the host
\citep[e.g.,][]{spi10}.\footnote{Note, however, that the detail of how
  the outflowing material interacts with the surroundings and perhaps
  regulates the inflow rate of gas is still subject to an active
  debate \citep{tas08,fel15}.} We determined a gas metallicity (12 +
log(O/H)) for each galaxy in our sample using the emission line method
developed in \cite{pet04} (using ([O {\sc iii}]/H$\beta$)/(N {\sc
  ii}/H$\alpha$)) with emission line strengths from a few sources (see
Table~\ref{tbl:sample}). In order to compare these gas metallicities
with the mass-metallicity relation derived in \cite{tre04}, we used
the metallicity calibration conversion found in \cite{kew08} to
calculate analogous values, taking into account the $\sim$0.1 dex
uncertainty in the \cite{pet04} method as well as the $\sim$0.06 dex
uncertainty in the conversion \citep{kew08}. The converted values are
listed in Table~\ref{tbl:sample}. The mass-metallicity relation
derived in \cite{tre04} is
$
12 + \text{log} (O/H)_{\rm predicted} \ = \ - \ 1.492 +
1.847(\text{log} M_*) - 0.08026(\text{log} M_*)^2,
$
%
where O/H is the oxygen abundance and $M_*$ is the stellar
mass in units of solar masses. The equation is valid over the mass
range 8.5 \textless \ log ($M_*/M_\odot$) \textless \ 11.5. We used this
relation and the $M_*$ values listed in Table~\ref{tbl:sample} to
calculate the predicted value of 12 + log(O/H)$_{\rm predicted}$. We have taken
solar metallicity to be 12 + log(O/H) = 8.69 \citep{all01}.

In Figure~\ref{fig:cgmass_delZ}, we calculate a metallicity deficit
($\Delta$ log(O/H)) by subtracting 12 + log(O/H)$_{\rm predicted}$
from the value derived from the emission line strengths and compare
that to $M_{cg}$/$M_{global}$. In NGC~3077 and He~2-10, dusty tidal
streams and AGN should be considered respectively when interpreting
Figure~\ref{fig:cgmass_delZ}. The tidal streams of NGC~3077 likely
increase $M_{cg}$/$M_{global}$, even though our observations do not
cover the entire spatial extent of the streams when compared with the
H {\sc i} data \citep{cot76,van79,yun94,wal11}. The AGN in He~2-10
\citep{rei11,rei16} may eject and heat dust more efficiently, thus
increasing $M_{cg}$/$M_{global}$. Also, the presence of the AGN may
raise [O {\sc iii}]/H$\beta$ resulting in an underestimate of the
actual log(O/H) value for this object. Keeping these considerations
for He~2-10 and NGC~3077 in mind, we find that the galaxy with the
largest metallicity deficit, NGC~1569, also has the largest fraction
of circumgalactic dust. In this context, it is interesting to note
that the amount of metals locked in the cold dust outflow of NGC~1569
we found with our MBB fits (4.5 $\times$ 10$^5$ $M_{\odot}$, assuming
most of the dust mass is in metals) is $\sim$ 10 times larger than the
mass of oxygen detected in its hot outflow \citep[3.4 $\times$ 10$^4$
  $M_{\odot}$;][]{mar02}.

Overall, the small number of objects in our sample and the rather
large uncertainties on $\Delta$ log(O/H) and $M_{cg}$/$M_{global}$
preclude us from drawing any statistically significant
conclusions. Furthermore, recent results have warned of the increased
scatter in the mass-metallicity relation at lower masses
\citep[e.g.,][]{zah12,paa17}, which may further cloud the
issue. Finally, as noted previously, the extent of the H {\sc i} in
dwarf galaxies is known to differ significantly from the stellar
component. Therefore, using the stellar component to define an edge or
border for the galaxy remains an imperfect method of determining what
is ``inside'' versus ``outside'' of the galaxy. This is clearly an
issue in NGC~3077 and also perhaps in NGC~1569
\citep[e.g.,][]{joh12,joh13}. In addition, the lack of velocity
information in our data makes it challenging to attribute
circumgalactic dust features to the known winds. Attempts to link the
cold dust features detected in the {\em Herschel} maps with other dust
tracers with kinematic information such as infrared H$_2$ and mm-wave
CO transitions, Na I~D 5890, 5896 \AA, and H {\sc i} 21 cm may provide
a way around the lack of direct kinematic constraints on the cold dust
component and a more concrete link to galactic winds \citep[e.g.,][and
  references therein]{rup13a,rup13b,bei15,ler15,rup17}.

\clearpage


\section{SUMMARY}
\label{summary}

In this second paper of a series, deep {\em Herschel Space
  Observatory} data of six nearby dwarf galaxies selected to host
known galactic-scale winds are used to investigate the structure and
properties of the cold dust in and around these systems. Our analysis
of these data has yielded the following results:

\begin{itemize}
\item 
  The gain in both angular resolution and sensitivity of {\em
    Herschel} PACS and SPIRE over {\em Spitzer} MIPS has allowed us to
  detect and resolve previously unsuspected cold dust features in all
  six objects of our sample. These features have diverse morphologies:
  some are filamentary or clumpy while others span a large range of
  galactocentric angles. They extend out to 1 $-$ 3 kpc, sometimes
  well beyond the stellar component.  Some of these features coincide
  spatially with emission features in H$\alpha$, 8.0, and 24 $\mu$m
  data, but others do not.

\item
  We have used archival {\em Spitzer} 4.5 $\mu$m IRAC data to estimate
  the spatial extent of the stellar components in these objects and
  distinguish the dust emission that coincides with the ``disk''
  region, defined by this stellar component, from the emission that
  comes from the ``circumgalactic'' region. We find that typically
  $\sim$10-20\% of the total dust mass resides in the circumgalactic
  region, reaching a value as large as $\sim$60\% in the case of
  NGC~1569. The presence of circumgalactic cold dust features in these
  dwarf galaxies does not obviously depend on the star formation rate
  surface density or gravitational influence of neighboring galaxies,
  although the latter almost certainly plays a role in some objects
  (e.g., NGC~3077). The galaxy with the largest metallicity (O/H)
  deficit in our sample, NGC~1569, also happens to have the largest
  fraction of circumgalactic dust. This may be coincidental. The small
  number of galaxies in our sample and the substantial uncertainties
  on the measurements of the dust mass and gas metallicity preclude us
  from drawing any solid conclusion regarding a possible physical link
  between metallicity and circumgalactic dust mass fraction.

\end{itemize}

Our data do not allow us to determine unambiguously the origin of the
circumgalactic dust. The best case in our sample for dust entrained in
a galactic wind is NGC~1569, but we cannot rule out the possibility
that cold dust features that appear to lie in the zone of influence of
the galactic wind are instead dust features of external (e.g., tidal)
origin in front or behind the wind.  Future studies that not only
consider the distribution of the circumgalactic dust with respect to the
wind and tidal features but also the kinematics of dust tracers such
H$_2$, CO, Na I~D, and H {\sc i} will help shed light on the origin of this
far-flung dust in NGC~1569 and other dwarf galaxies. 

\vskip 0.25in

\section*{ACKNOWLEDGEMENTS}

This work is based in part on observations made with {\em Herschel},
which is an ESA space observatory with science instruments provided by
European-led Principle Investigator consortia and with important
participation from NASA. Support for this work was provided by NASA
through {\em Herschel} contracts 1427277 and 1454738 (AM, SV, and MM)
and ADAP grant NNX16AF24G (SV). Support for this work was also
provided by the National Science Foundation under AST-1109288
(CLM). Support for this work was also provided by European Union's
Horizon 2020 Research and Innovation Programme, under Grant
Agreement number 687378 (TM). We acknowledge the helpful
comments of the referee. We thank Alberto Bolatto, Stuart Vogel,
and Janice C. Lee for helpful comments and suggestions during the
preparation of the manuscript. We thank H\'el\`ene Roussel for her guidance
in using the {\em Scanamorphos} software and Peter Teuben for his help
converting old H$\alpha$ data. This work has made use of NASAs
Astrophysics Data System Abstract Service and the NASA/IPAC
Extragalactic Database (NED), which is operated by the Jet Propulsion
Laboratory, California Institute of Technology, under contract with the
National Aeronautics and Space Administration. PACS has been developed
by a consortium of institutes led by MPE (Germany) and including UVIE
(Austria); KU Leuven, CSL, IMEC (Belgium); CEA, LAM (France); MPIA
(Germany); INAF-IFSI/OAA/OAP/OAT, LENS, SISSA (Italy); IAC (Spain).
This development has been supported by the funding agencies BMVIT
(Austria), ESA-PRODEX (Belgium), CEA/CNES (France), DLR (Germany),
ASI/INAF (Italy), and CICYT/MCYT (Spain). SPIRE has been developed by
a consortium of institutes led by Cardiff University (UK) and
including Univ. Lethbridge (Canada); NAOC (China); CEA, LAM (France);
IFSI, Univ. Padua (Italy); IAC (Spain); Stockholm Observatory
(Sweden); Imperial College London, RAL, UCL-MSSL, UKATC, Univ. Sussex
(UK); and Caltech, JPL, NHSC, Univ. Colorado (USA). This development
has been supported by national funding agencies: CSA (Canada); NAOC
(China); CEA, CNES, CNRS (France); ASI (Italy); MCINN (Spain); SNSB
(Sweden); STFC, UKSA (UK); and NASA (USA). HIPE is a joint development
(are joint developments) by the Herschel Science Ground Segment
Consortium, consisting of ESA, the NASA Herschel Science Center, and
the HIFI, PACS and SPIRE consortia.

\newpage
\appendix
\section{NOTES ON INDIVIDUAL GALAXIES}
\label{appendix}

For the discussion in these notes, please refer to
Figures~\ref{fig:He2-10} -~\ref{fig:n5253} and
Figures~\ref{fig:He2-10_CLEAN} -~\ref{fig:n5253_CLEAN}.

\begin{itemize}
\item {\em He~2-10} - Broad cold dust emission regions surround
  He~2-10 out to $\sim$ 1.5 kpc in all SPIRE bands as well as the 160
  $\mu$m PACS map. The 160 $\mu$m emission map appears similar in extent
  and morphology to the H {\sc i} distribution observed by
  \cite{kob95}. The 70 $\mu$m PACS map exhibits a bit more anisotropy
  in its distribution of circumgalactic emission with tapering
  features extending to the north-northeast, southwest, and
  east-southeast. The north-northeast feature extends furthest to
  $\sim$ 2.6 kpc. The orientation of these tapering features appears
  similar to the smaller 0.5 kpc-scale H$\alpha$ bubbles
  \citep{men99,zas13}, but appears less correlated with the shape of
  the X-ray emission \citep{kob10}. Several knots of emission
  separated from He~2-10 by $\sim$3-4 kpc to the northwest and
  southwest in the PACS maps cannot be definitively ruled out as
  background sources given the signal-to-noise levels or foreground
  Milky Way dust clouds, but coincident emission in the SPIRE maps
  seems connected to the disk via colder streams bridging their
  separation. The most prominent emission in the SPIRE maps towards
  the knots northwest of the disk seems similarly oriented to an X-ray
  feature observed by \cite{kob10}.

\item {\em NGC~1569} - A dense complex of filamentary and broad
  circumgalactic cold dust features stretch out from all along the
  disk of NGC~1569. The prominent H$\alpha$, 8.0, and 24 $\mu$m
  filament \citep{wal91,hun93,wes08,mcc13} extending southwest from
  the western edge of the disk also shows up in the PACS maps, but
  becomes fainter and disappears in the SPIRE bands. A clump of cold
  dust directly south of the galactic nucleus resides $\sim$1.8 kpc
  from the disk midplane and appears to have made a clean break with
  the other circumgalactic cold dust emission in the PACS maps. Some
  faint 8.0 $\mu$m emission \citep{mcc13} coincides with this rogue
  clump and some extended X-ray emission \citep{hec95,del96} may as
  well. The rogue clump has no obvious counterpart in H$\alpha$
  emission, but supershell features surrounding its location
  \citep{mar98,wes08} suggest that it resides at the center of an
  ionized gas shell. The structure of the rogue clump suggests
  anisotropic heating with the warmer 70 $\mu$m emission closer to the
  galactic disk and the colder emission in the SPIRE maps extending
  further away. The bright emission to the southwest corner of the
  SPIRE maps comes from foreground Milky Way cirrus, although emission
  from dust possibly associated with the large-scale H~I complex
  studied by \citep{joh12,joh13} cannot be ruled out. Note that some
  dispute remains about the distance to NGC~1569 \citep{gro08}. We
  have adopted the distance determined by \cite{gro08}, which employs
  {\em Hubble Space Telescope} data and the tip of the red giant
  branch to obtain a redshift-independent distance of 3.36 Mpc, but
  the range of published distance measurements (1.9 $-$ 3.4 Mpc) leads
  to a broad range of star formation rates.

\item {\em NGC~1705} - Vertical filamentary cold dust emission to the
  north and south of NGC~1705 coincides with the H {\sc i} emission,
  which \cite{meu98} has shown consists mainly of a rotating
  disk. However, Meurer et al. also note an H {\sc i} ``spur" which is
  kinematically distinct from the rotating disk and spatially
  consistent with the H$\alpha$ outflow \citep{meu89,meu92,zas13}. Our
  PACS maps show evidence of a similarly consistent cold dust feature
  extending almost exactly north from the galaxy. Also in the PACS
  maps, two bright knots of emission directly to the south are
  separated from the galactic center by $\sim$1.3 and 1.5 kpc,
  respectively. These knots are also approximately coincident with the
  rotating H {\sc i} disk, also appear in the SPIRE maps, and do not
  have associated background sources. North-northeast of the nucleus,
  a bright cloud of cold dust appears as a knot in the 70 $\mu$m PACS
  map but has broader spatial extent in our other four {\em Herschel}
  maps. This cloud with no obviously associated background source
  seems spatially consistent with the orientation of the northern
  H$\alpha$ superbubble \citep{zas13}, but resides well beyond its
  edge (\textgreater 0.5 kpc), making it a likely example of material
  swept out of the ISM by an expanding superbubble. However, we did
  not include this cloud in the global flux measurements, since it
  cannot be definitively ruled out as a background source.

\item {\em NGC~1800} - A C-shaped filament of circumgalactic cold
  dust extends from the northwestern edge of NGC~1800 to $\sim$1.2 kpc
  north of the disk in the 70 $\mu$m PACS map tracing a similar
  feature in 8.0 $\mu$m emission. In addition there is some more
  diffuse, less bright 70 $\mu$m emission above the disk to the east
  of this filament. These features lie just south of the filamentary
  web of H$\alpha$ emission \citep{hun94,mar95,hun96}. The cold dust
  features along the northwestern edge of NGC~1800 are offset from the
  H$\alpha$ ``fingers" of \cite{hun96}, which extend from the
  northeastern edge of the disk. An offset also exists between the
  southern edge cold dust features which appear like two broad nubs
  splayed out from the eastern and western edges of the disk in our
  SPIRE maps versus the centrally concentrated H$\alpha$ ``fingers"
  and shell outside the disk (again from
  \citealt{hun96}). Observations by \cite{ras04} reveal X-ray emission
  crossing the disk of NGC~1800 approximately from southeast to
  northwest at an angle with respect to the stellar component. The
  northern extent of this hot X-ray gas above the disk appears to
  coincide with some of the emission from cold dust in the PACS
  maps. Within the galactic disk, the peak of the dust emission is
  offset to the west from the stellar component, perhaps helping
  explain why the strongest dust feature $-$ the C-shaped filament
  extending north $-$ emanates from that part of galaxy. This offset
  between dust and stellar emission might be explained by ram pressure
  stripping, but the galaxy appears to be relatively isolated
  \citep{kar14}. What may appear as cold dust knots to the south in
  the PACS images could also be either Galactic dust heated by
  foreground Milky Way stars or emission from background sources
  evident in the 4.5 $\mu$m IRAC map.

\item {\em NGC~3077} - Most if not all of the circumgalactic cold dust
  features we observe for NGC~3077 can be attributed to the same tidal
  interaction forces with M81 and M82, which produce tidal arms of H
  {\sc i} and the complex of H {\sc i} and CO offset to the east of
  the galaxy's stellar component
  \citep{cot76,van79,yun94,wal99}. However, certain cold dust features
  appear to complement the morphology of the H$\alpha$ emission
  \citep{mar98,cal04}. A C-shaped filament and a knot of cold dust at
  its tip extending $\sim$2.6 kpc north of the stellar disk (see 70 \&
  160 $\mu$m maps) reside directly above the location of an H$\alpha$
  shell \citep{cal04}. A circumferential arc of cold dust visible in
  the PACS maps and the 250 $\mu$m SPIRE map to the south of the disk
  also traces the edge of a superbubble observed in H$\alpha$ and
  X-ray emission \citep{mar98,ott03}. Cold dust filaments, clouds,
  broad regions, and clumps to the east and west of NGC~3077 are
  likely associated with the H {\sc i} features
  \citep{yun94,wal11}. The argument has been made that the
  superbubbles of hot plasma and ionized gas have yet to break out of
  this galaxy but potentially have enough energy to generate a
  breakout wind in the northern direction where there is less H {\sc
    i} density \citep{ott03}. The cold dust emission along the
  north-south direction in the PACS maps appears stronger than the
  east-west emission while the inverse is true of the east-west
  emission in the SPIRE maps, which would seem to support the argument
  for a wind developing in the northern direction. Our deep PACS data
  provide evidence for this north-south emission, which was not
  identified in a previous analysis of the SPIRE data \citep{wal11}.

\item {\em NGC~5253} - The morphology of the circumgalactic cold dust
  features in our PACS and SPIRE maps appears to follow the
  distribution of H {\sc i} \citep{kob08} more closely than any other
  galaxy component. However, features like the broad emission regions
  extending from the northwest and southeast edges of the stellar disk
  in the {\em Herschel} maps do follow ionized gas features
  \citep{mar95,cal99,zas11}, X-ray emission \citep{m+k95,str99}, and
  some 8.0 $\mu$m features \citep{mcc13}. Filaments extend south in
  the 70, 160, and 250 $\mu$m maps, where the westernmost filament
  protruding from the southwest edge of the disk extends at least
  $\sim$2 kpc from the disk. In the PACS maps, a circumgalactic cold
  dust filament also extends directly north from the disk. Looking
  along the north-south direction, the 500 $\mu$m SPIRE map shows
  tenuous evidence for a significantly larger scale bipolar
  filamentary emission extending to $\sim$5 kpc outside the galactic
  disk. The southern extent of this emission appears disturbed and
  perhaps coincides with the H {\sc i} plume interpreted by
  \cite{kob08} as a potential outflow or inflow. However, no obvious
  counterpart for the northern filament presents itself in the H {\sc
    i} images. The challenge of ruling out background sources or cold
  Milky Way dust clouds remains a concern for this north-south
  emission, but the brightest features from apparent clouds within
  this structure are \textgreater 4$\sigma$ above the background in
  the 500 $\mu$m SPIRE map.

\end{itemize}

\newpage


\newpage
\normalsize


\noindent {\em Note: Figure resolution reduced to keep PDF file size under 10MB. Full resolution figures availabe in the published verison.}

\renewcommand{\thefigure}{\arabic{figure}\alph{subfigure}}
\setcounter{subfigure}{1}

\begin{figure}[htbp]
\centering
\includegraphics[width=1.0\textwidth]{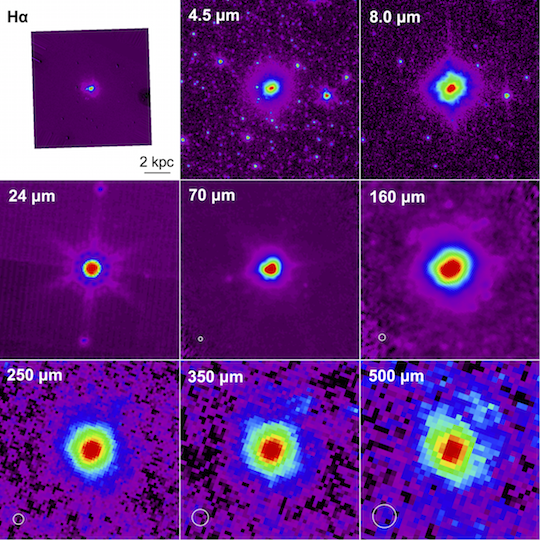}
\caption{\footnotesize{He~2-10 - H$\alpha$ \citep{zas13}, IRAC 4.5 and 8.0 $\mu$m and MIPS 24 $\mu$m ({\em Spitzer} archive, this work), PACS 70 and 160 $\mu$m, and SPIRE 250, 350, and 500 $\mu$m (this work). All images are displayed on a logarithmic intensity scale. North is up and east is to the left in all images. The bar in the H$\alpha$ image indicates the scale in all images. The white circles in the 70, 160, 250, 350, and 500 $\mu$m panels indicate the full width half maximum of the point spread function for each channel.}}
\label{fig:He2-10}
\end{figure}

\addtocounter{figure}{-1}
\addtocounter{subfigure}{1}
\begin{figure}[htbp]
\centering
\includegraphics[width=1.0\textwidth]{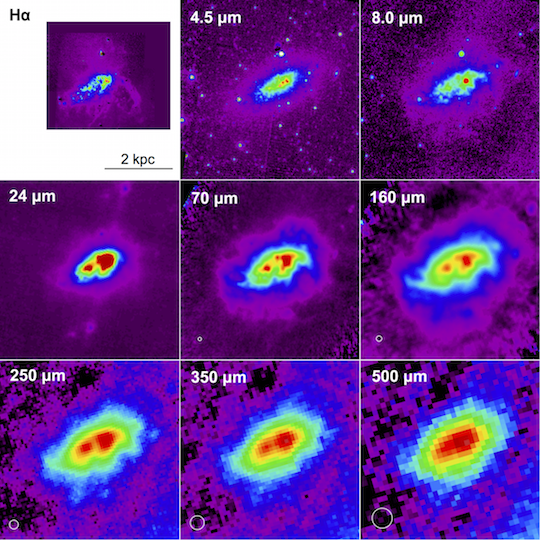}
\caption{\footnotesize{NGC~1569 - H$\alpha$ \citep{mar97}, IRAC 4.5 and 8.0 $\mu$m \citep{mcc13}, otherwise, the same as in Figure~\ref{fig:He2-10}.}}
\label{fig:n1569}
\end{figure}

\addtocounter{figure}{-1}
\addtocounter{subfigure}{1}
\begin{figure}[htbp]
\centering
\includegraphics[width=1.0\textwidth]{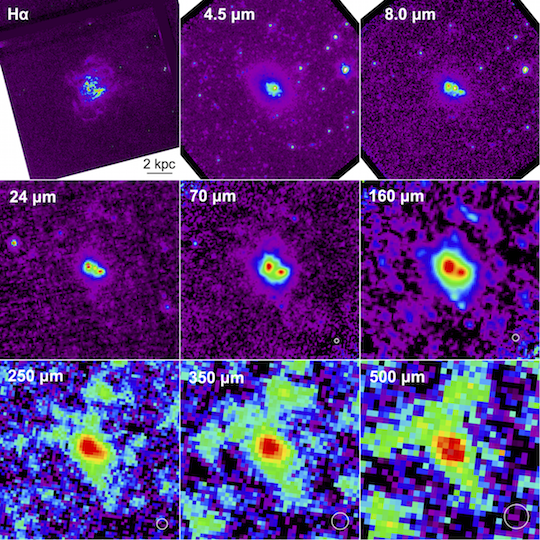}
\caption{\footnotesize{NGC~1705 - H$\alpha$ \citep{zas13}, IRAC 4.5 and 8.0 $\mu$m \citep{mcc13}, otherwise, the same as in Figure~\ref{fig:He2-10}.}}
\label{fig:n1705}
\end{figure}

\addtocounter{figure}{-1}
\addtocounter{subfigure}{1}
\begin{figure}[htbp]
\centering
\includegraphics[width=1.0\textwidth]{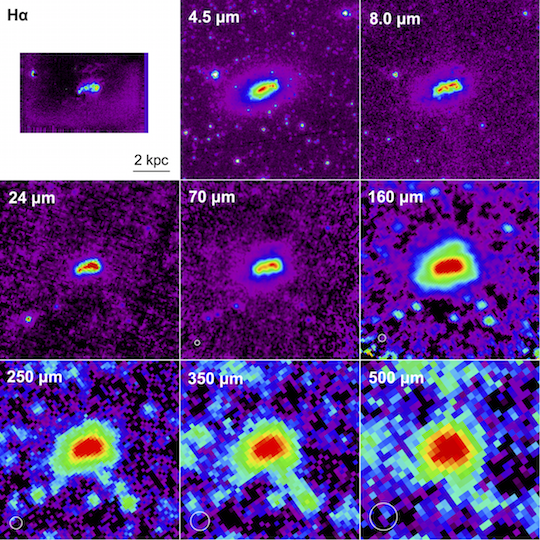}
\caption{\footnotesize{NGC~1800 - H$\alpha$ \citep{mar97}, otherwise, the same as in Figure~\ref{fig:He2-10}.}}
\label{fig:n1800}
\end{figure}

\addtocounter{figure}{-1}
\addtocounter{subfigure}{1}
\begin{figure}[htbp]
\centering
\includegraphics[width=1.0\textwidth]{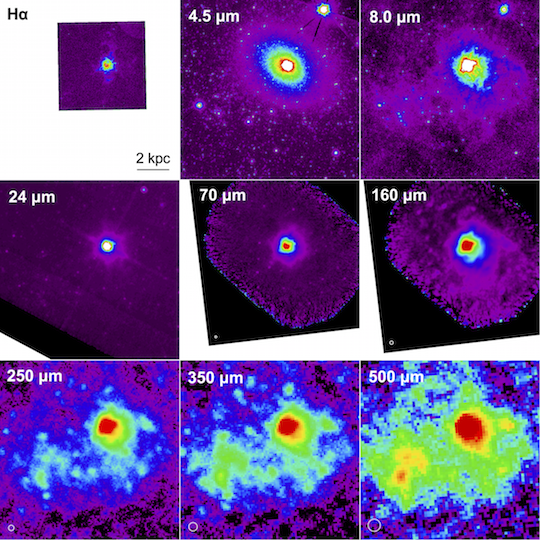}
\caption{\footnotesize{NGC~3077 - H$\alpha$ \citep{dale09}, otherwise, the same as in Figure~\ref{fig:He2-10}.}}
\label{fig:n3077}
\end{figure}

\addtocounter{figure}{-1}
\addtocounter{subfigure}{1}
\begin{figure}[htbp]
\centering
\includegraphics[width=1.0\textwidth]{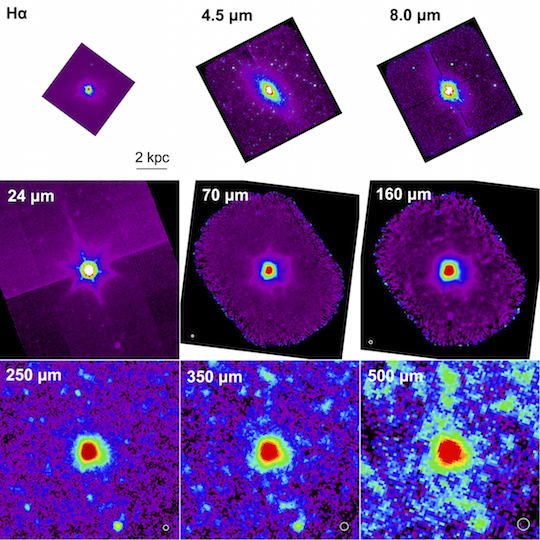}
\caption{\footnotesize{NGC~5253 - H$\alpha$ \citep{zas11}, IRAC 4.5 and 8.0 $\mu$m \citep{mcc13}, otherwise, the same as in Figure~\ref{fig:He2-10}.}}
\label{fig:n5253}
\end{figure}

\renewcommand{\thefigure}{\arabic{figure}}

\FloatBarrier

\renewcommand{\thefigure}{\arabic{figure}\alph{subfigure}}
\setcounter{subfigure}{1}

\begin{figure}[htbp]
\centering
\includegraphics[width=1.0\textwidth]{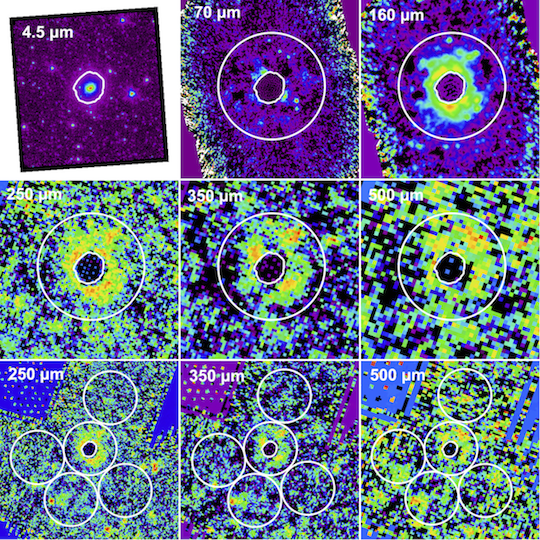}
\caption{\footnotesize{{\em Herschel} maps of He~2-10 at 70, 160, 250,
    350, and 500 $\mu$m after application of our CLEAN algorithm. The
    IRAC 4.5 $\mu$m map shows the stellar component for
    comparison. All images are displayed on a logarithmic intensity
    scale. North is up and east is to the left in all images. The
    inner white contour in the {\em Herschel} maps indicates the
    stellar disk region derived from the 4.5 $\mu$m data, while the
    white circle or ellipse shows the full extent over which the
    global infrared fluxes were calculated. The bottom row shows the
    SPIRE maps (250, 350, and 500 $\mu$m) and the circular or
    elliptical regions used to estimate the background flux in each
    band. The field of view of the panels in the bottom row is twice
    that of the six panels in the top and middle rows.}}
\label{fig:He2-10_CLEAN}
\end{figure}

\addtocounter{figure}{-1}
\addtocounter{subfigure}{1}
\begin{figure}[htbp]
\centering
\includegraphics[width=1.0\textwidth]{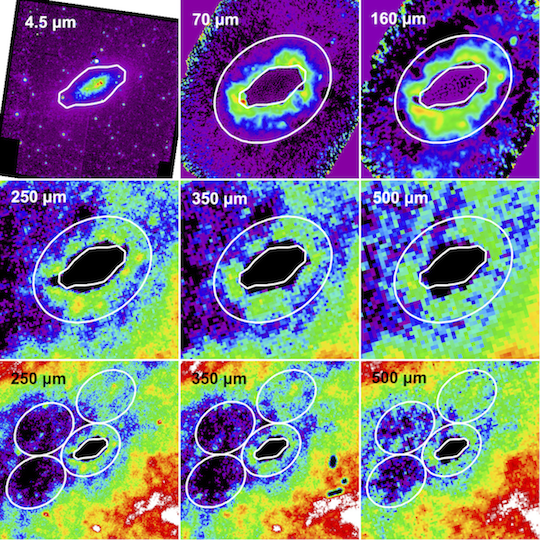}
\caption{\footnotesize{NGC~1569 - display is the same as in Figure~\ref{fig:He2-10_CLEAN}}}
\label{fig:n1569_CLEAN}
\end{figure}

\addtocounter{figure}{-1}
\addtocounter{subfigure}{1}
\begin{figure}[htbp]
\centering
\includegraphics[width=1.0\textwidth]{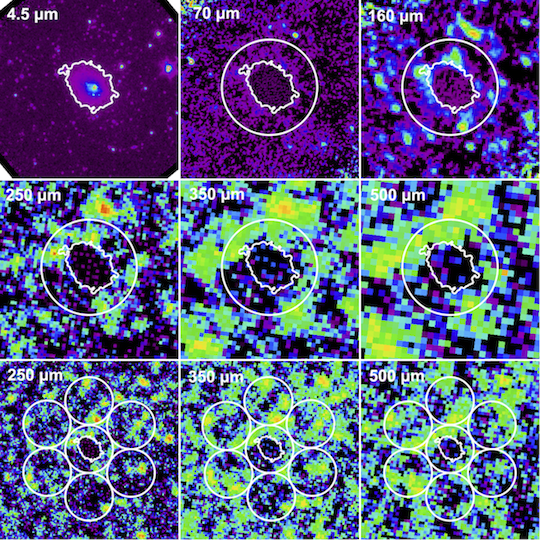}
\caption{\footnotesize{NGC~1705 - display is the same as in Figure~\ref{fig:He2-10_CLEAN}}}
\label{fig:n1705_CLEAN}
\end{figure}

\addtocounter{figure}{-1}
\addtocounter{subfigure}{1}
\begin{figure}[htbp]
\centering
\includegraphics[width=1.0\textwidth]{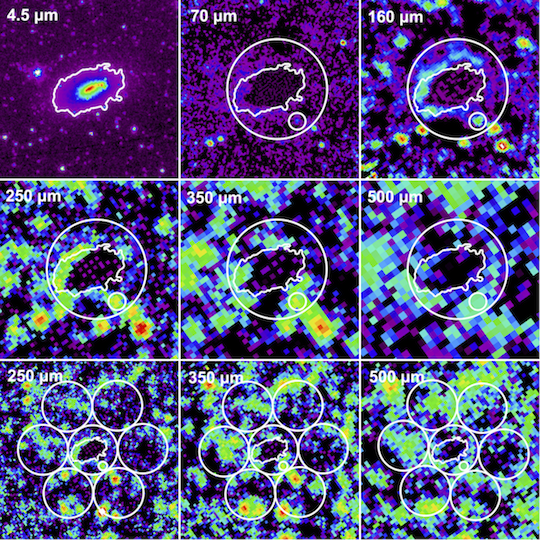}
\caption{\footnotesize{NGC~1800 - display is the same as in Figure~\ref{fig:He2-10_CLEAN}, plus one region is masked (smaller white circle).}}
\label{fig:n1800_CLEAN}
\end{figure}

\addtocounter{figure}{-1}
\addtocounter{subfigure}{1}
\begin{figure}[htbp]
\centering
\includegraphics[width=1.0\textwidth]{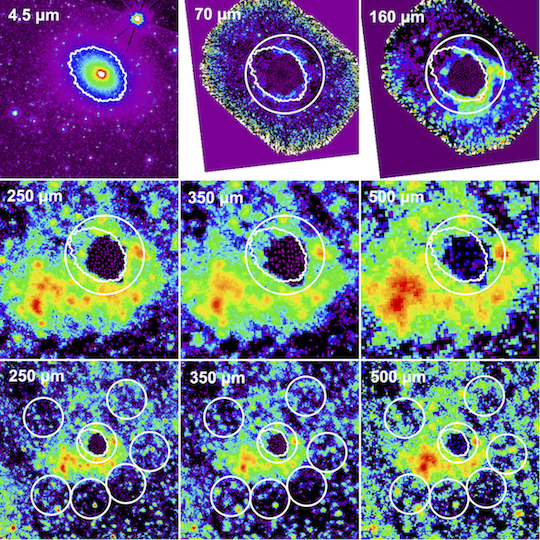}
\caption{\footnotesize{NGC~3077 - display is the same as in Figure~\ref{fig:He2-10_CLEAN}}}
\label{fig:n3077_CLEAN}
\end{figure}

\addtocounter{figure}{-1}
\addtocounter{subfigure}{1}
\begin{figure}[htbp]
\centering
\includegraphics[width=1.0\textwidth]{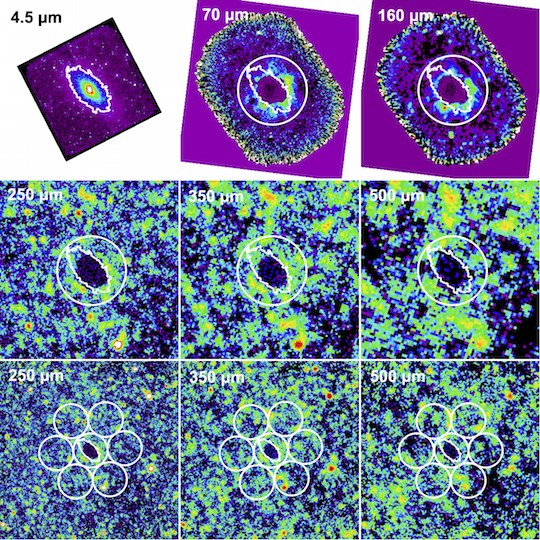}
\caption{\footnotesize{NGC~5253 - display is the same as in Figure~\ref{fig:He2-10_CLEAN}}}
\label{fig:n5253_CLEAN}
\end{figure}

\renewcommand{\thefigure}{\arabic{figure}}

\FloatBarrier

\begin{figure}[htbp]
\centering
\includegraphics[width=1.0\textwidth]{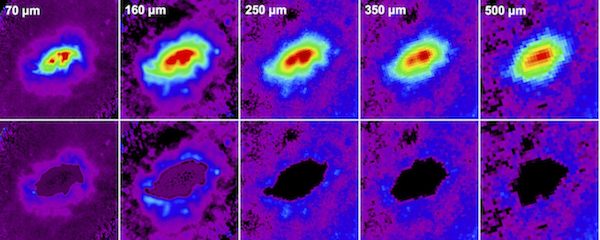}
\caption{\footnotesize{Top row shows the PACS and SPIRE maps of
    NGC~1569 while the bottom row shows the resulting residual maps
    after application of our modified CLEAN algorithm. We defined the
    regions where we applied the CLEAN algorithm in each band based on
    the stellar disk region for NGC~1569. Since each map has a
    different pixel scale and slightly different morphology to the
    bright disk areas, the regions where we applied the CLEAN
    algorithm are similar in extent but necessarily slightly different
    from band to band. All maps are shown on logarithmic intensity
    scale. North is up and east is to the left in all maps.}}
\label{fig:CLEAN}
\end{figure}

\FloatBarrier

\begin{figure}[htbp]
\centering
\includegraphics[width=0.6\textwidth]{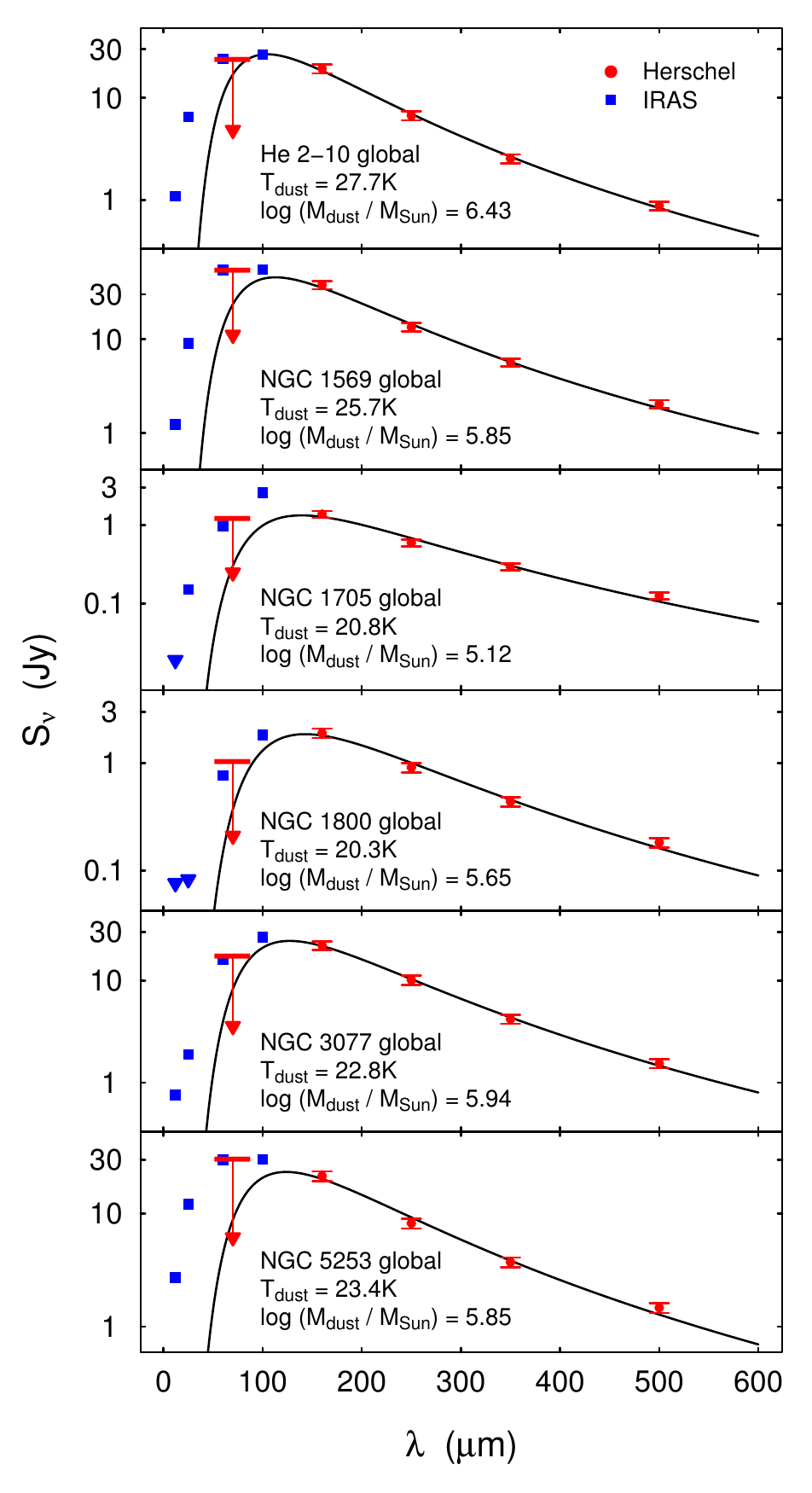}
\caption{\footnotesize{Modified blackbody fits (shown as black curves)
    to the global galaxy fluxes derived from the 70, 160, 250, 350,
    and 500 $\mu$m {\em Herschel} data (shown as red points). In this
    fitting, the 70 $\mu$m flux values were treated as upper limits as
    indicated by the downward arrows. The fit parameters are
    $M_{dust}$ and $T_{dust}$. Refer to \S~\ref{sed_fit} for
    details. The blue squares mark the {\em IRAS} fluxes at 12,
    25, 60, and 100 $\mu$m \citep{mos90,san03}, while the blue
    downward triangles represent upper limits. The modified blackbody
    fits do not include the {\em IRAS} flux values since these data
    are shallower than our {\em Herschel} maps.}}
\label{fig:mbb_global}
\end{figure}

\begin{figure}[htbp]
\centering
\includegraphics[width=0.6\textwidth]{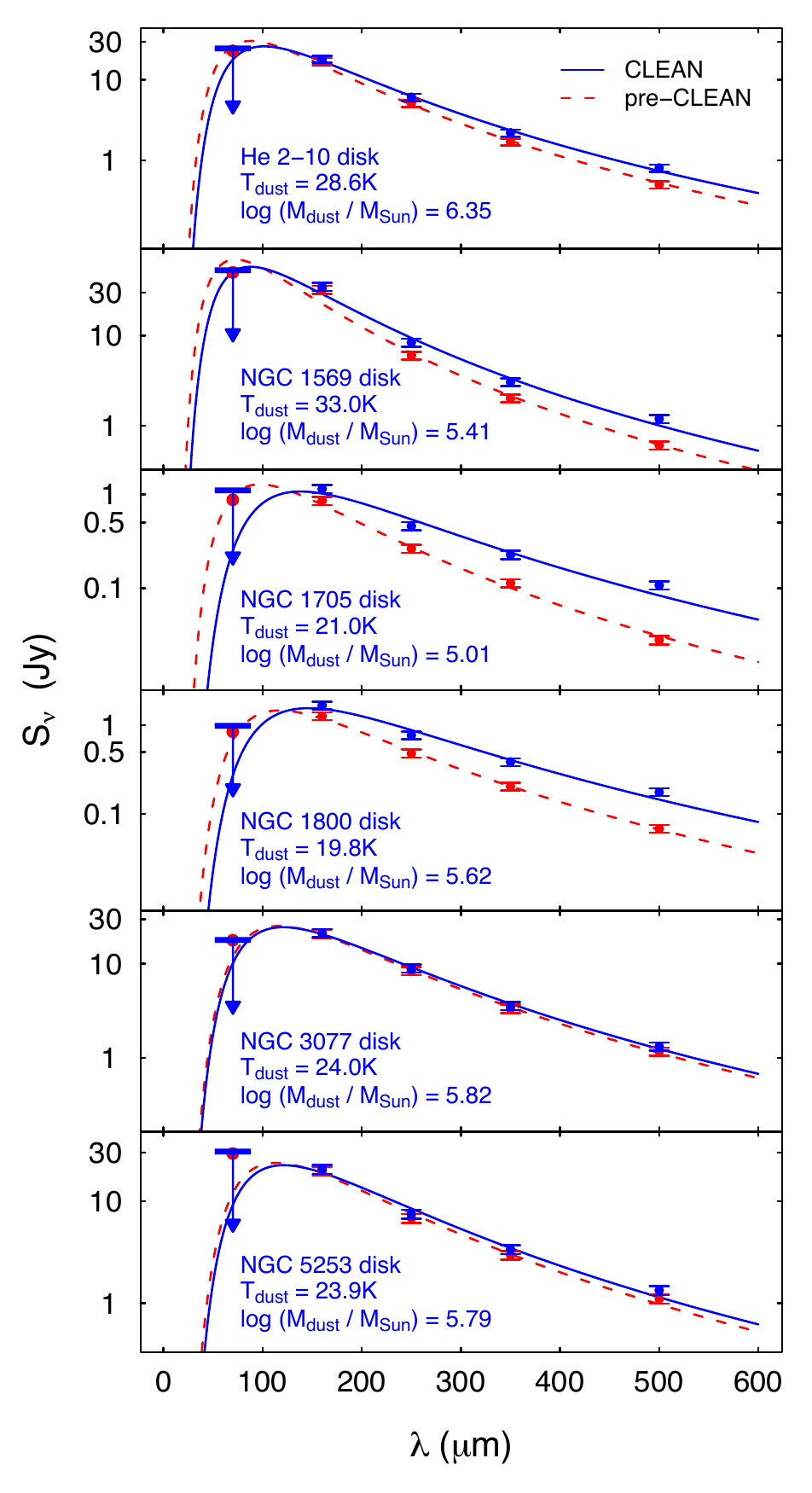}
\caption{\footnotesize{Modified blackbody fits to the galaxy disk
    fluxes derived from the 70, 160, 250, 350, and 500 $\mu$m {\em
      Herschel} maps before (dashed red line) and after (solid blue
    line) applying our modified CLEAN algorithm. In the fitting, the
    70 $\mu$m flux values were treated as upper limits as indicated by
    the downward arrows. The fit parameters are $M_{dust}$ and
    $T_{dust}$. Refer to \S~\ref{sed_fit} for details.}}
\label{fig:mbb_disk}
\end{figure}

\begin{figure}[htbp]
\centering
\includegraphics[width=1.0\textwidth]{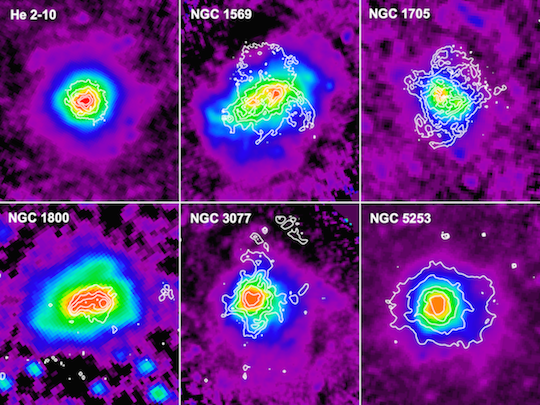}
\caption{\footnotesize{PACS 160 $\mu$m maps overlaid with H$\alpha$
    contours to compare the distribution of the cold dust with that of
    the warm ionized material. The field of view (FOV) of each panel is
    different to show the details at both 160 $\mu$m and H$\alpha$.
    FOV: He~2-10 3.96' x 4.46'; NGC~1569 4.79' x 5.40'; NGC~1705
    3.13' x 3.53'; NGC~1800 2.64' x 2.98'; NGC~3077 5.23' x 5.89';
    NGC~5253 4.01' x 4.52'.}}
\label{fig:halpha}
\end{figure}

 \begin{figure}[htbp]
\centering
\includegraphics[width=1.0\textwidth]{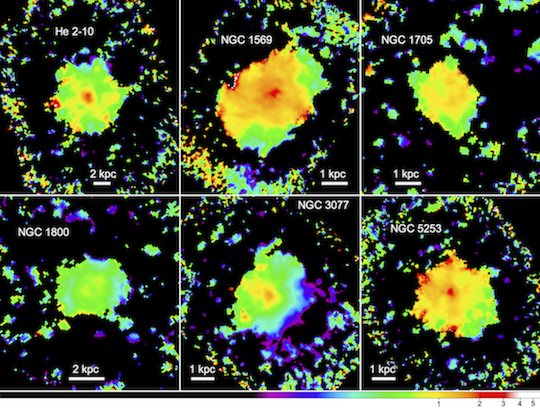}
\caption{\footnotesize{PACS 70$\mu$m / 160$\mu$m flux ratio maps. All
    maps are shown on the same logarithmic scale indicated by the
    color bar. The white bar in each panel indicates the scale of each
    image. North is up and east is to the left in all images.}}
\label{fig:ratios}
\end{figure}

\begin{figure}[htbp]
\centering
\includegraphics[width=1.0\textwidth]{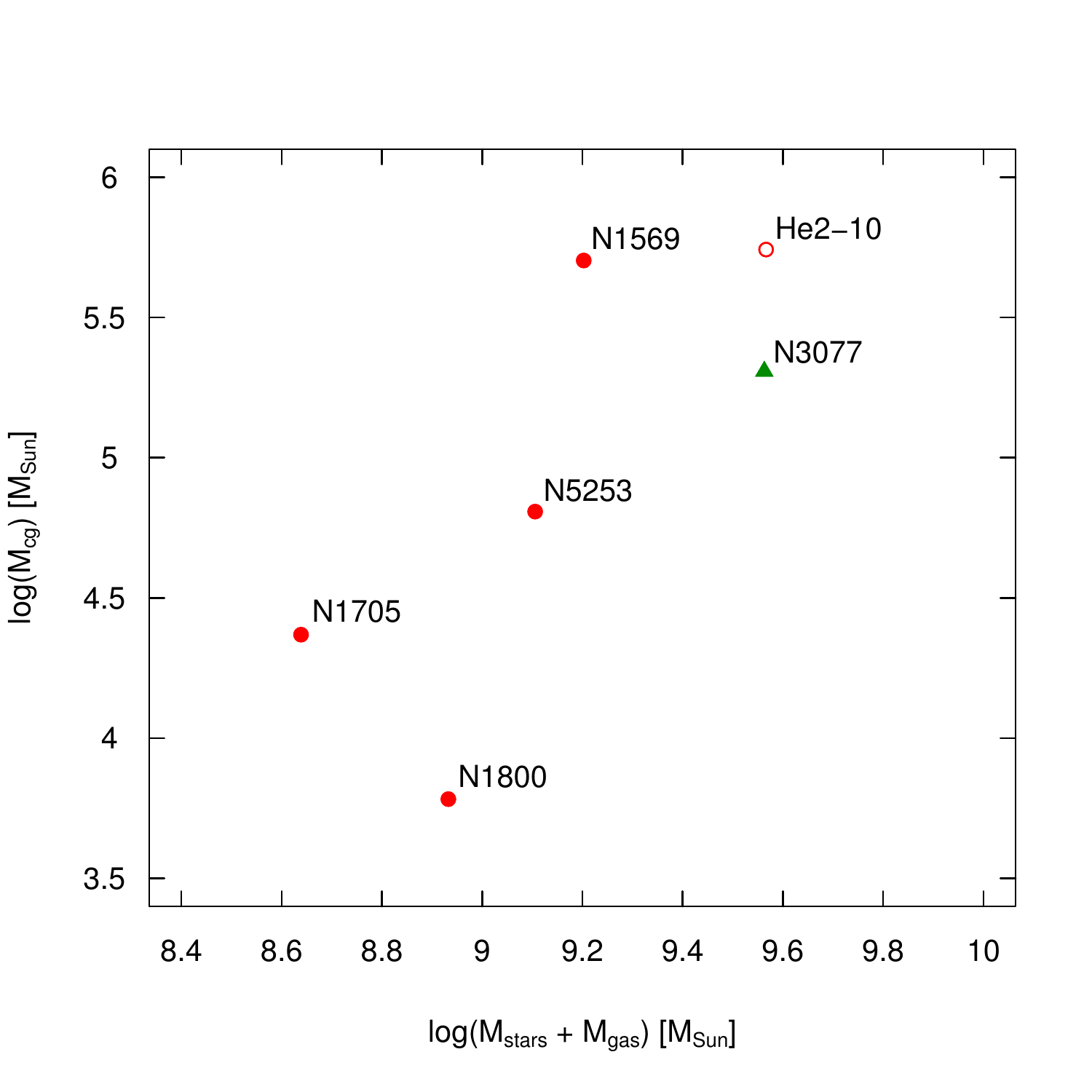}
\caption{\footnotesize{The circumgalactic dust mass is plotted against
    the total baryonic (stellar plus gas) mass, where the gas mass is
    the sum of the H {\sc i} and H$_2$ gas masses. NGC~3077 is
    differentiated with a green triangle to indicate the likely tidal
    contribution to its circumgalactic dust features, and He~2-10 is
    differentiated with an open circle to indicate the possible AGN
    contribution to the far-infrared flux and therefore derived dust
    mass. With those caveats, this plot suggests a weak (Pearson's $r$ =
    0.80; P[null] 0.055) positive correlation between circumgalactic
    dust mass and total baryonic mass.}}
\label{fig:cg_tot}
\end{figure}

\begin{figure}[htbp]
\centering
\includegraphics[width=1.0\textwidth]{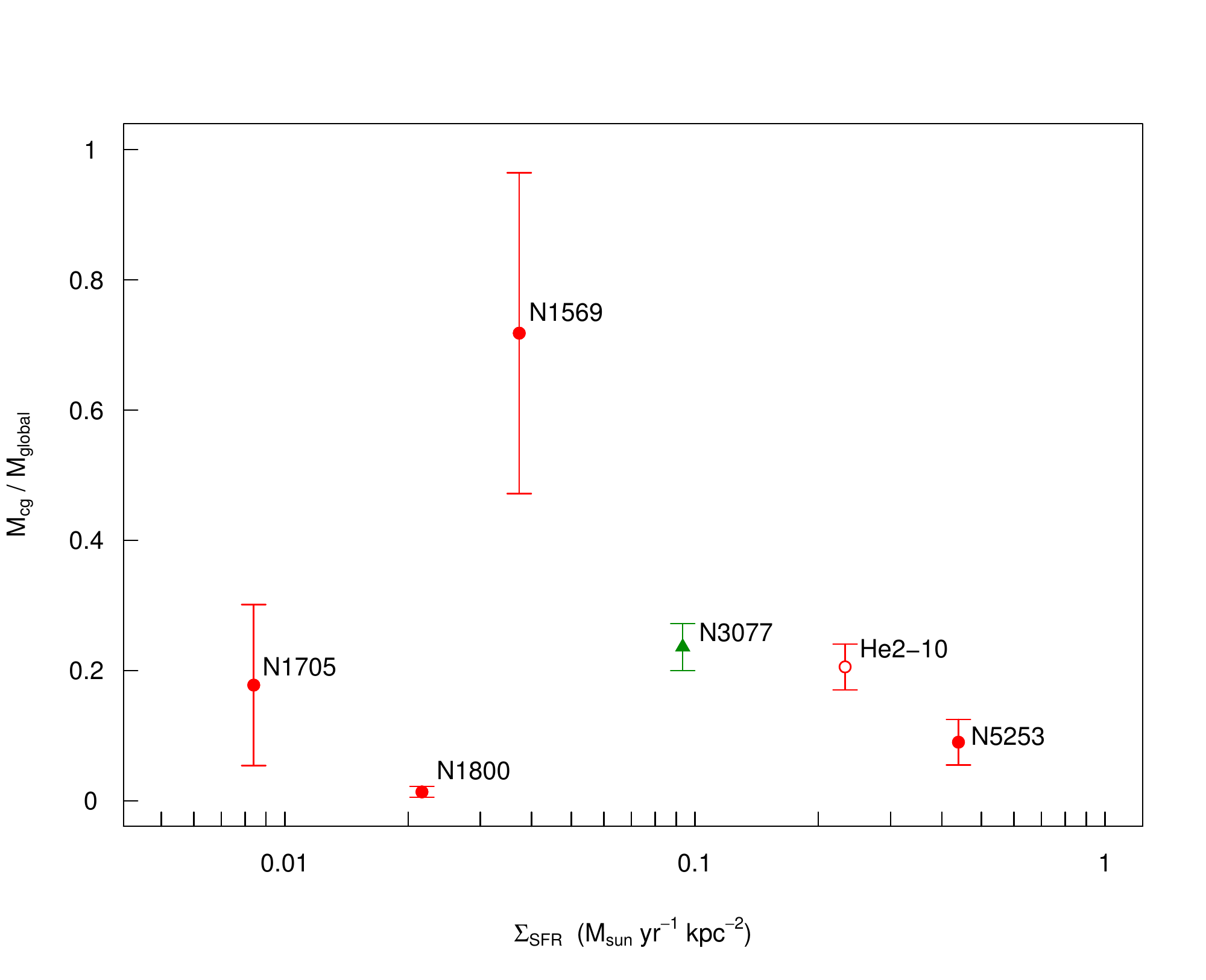}
\caption{\footnotesize{The circumgalactic dust mass fraction,
    $M_{cg}$/$M_{global}$, is plotted against the star formation rate
    surface density, $\Sigma_{SFR}$ from Table~\ref{tbl:sample}. The
    uncertainty in $M_{cg}$/$M_{global}$ comes from the uncertainties
    in the SED fitting (see \S~\ref{sed_fit} and
    Table~\ref{tbl:sed}). The meaning of the symbols is the same as in
    Figure~\ref{fig:cg_tot}. No trend is seen in this
    figure between $M_{cg}$/$M_{global}$ and $\Sigma_{SFR}$. }}
\label{fig:cgmass_Sigma}
\end{figure}

\begin{figure}[htbp]
\centering
\includegraphics[width=1.0\textwidth]{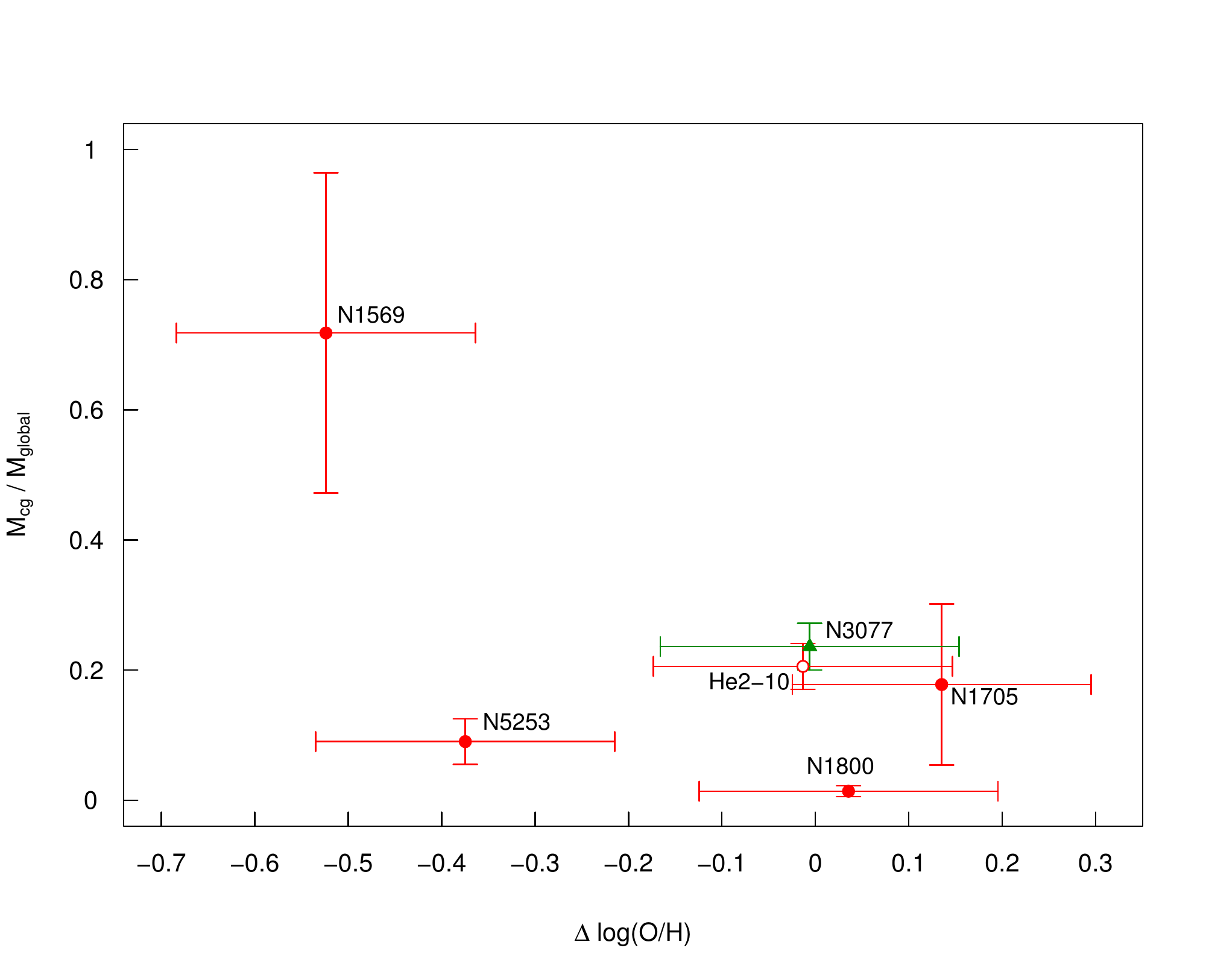}
\caption{\footnotesize{The circumgalactic dust mass fraction,
    $M_{cg}$/$M_{global}$, is plotted against the metallicity deficit,
    $\Delta$ log(O/H), defined as the difference between the galaxy
    metallicity (see Table~\ref{tbl:sample}) and the metallicity
    predicted by the relation derived in \cite{tre04} between $M_*$
    and 12 + log(O/H). The uncertainty on $\Delta$ log(O/H) comes from
    the uncertainty in the galaxy metallicity ($\sim$0.1 dex) and the
    uncertainty in converting values to conform with \cite{tre04}
    ($\sim$0.06 dex). The meaning of the symbols is the same as in
    Figure~\ref{fig:cg_tot}. No significant trend is seen in this
    figure between $M_{cg}$/$M_{global}$ and $\Delta$ log(O/H),
    although the object (NGC~1569) with the largest circumgalactic
    dust mass fraction also has the largest metallicity deficit. }}
\label{fig:cgmass_delZ}
\end{figure}

\FloatBarrier


\begin{deluxetable}{l l l c c c l c l l l l l l}
\tablecaption{Nearby Dwarf Galaxies and Their Properties \label{tbl:sample}}
\tabletypesize{\tiny}
\rotate
\tablecolumns{10}
\tablewidth{0pt}
\tablehead{
	\colhead{Galaxy} & 
	\colhead{Type} & 
	\colhead{Morph.} &
	\colhead{$D_{25}$} & 
	\colhead{$d$} &
	\colhead{scale} &
	\colhead{$L_{IR}$} &
	\colhead{12 +}&
	\colhead{$M_{HI}$} & 
	\colhead{$M_{H_2}$} &
	\colhead{log($M_*$/$M_{\odot}$)} &
	\colhead{SFR} &
	\colhead{$\Sigma_{SFR}$} &
	\colhead{Wind/eDIG Ref.} \\
	\colhead{} &
	\colhead{} &
	\colhead{} &
	\colhead{(')} &
	\colhead{(Mpc)} &
	\colhead{(pc/")} &
	\colhead{(10$^{9}$$L_{\odot}$)} &	
	\colhead{log(O/H)} &
	\colhead{10$^8$$M_{\odot}$} &
	\colhead{10$^8$$M_{\odot}$} &
	\colhead{} &
	\colhead{$M_{\odot}$ yr$^{-1}$} &
	\colhead{$M_{\odot}$ yr$^{-1}$ kpc$^{-2}$} &
	\colhead{} \\
	\colhead{[1]} &
	\colhead{[2]} &
	\colhead{[3]} &
	\colhead{[4]} &
	\colhead{[5]} &
	\colhead{[6]} &
	\colhead{[7]} &
	\colhead{[8]} &
	\colhead{[9]} &
	\colhead{[10]} &
	\colhead{[11]} &
	\colhead{[12]} &
	\colhead{[13]} &
	\colhead{[14]}
}
\startdata
He~2-10      & H {\sc ii} + AGN	& I0? pec			& 1.74  & 10.5	& 50.9  & 7.00		& 8.80  & 3.37  & 1.85    & 9.5$^{\dagger}$ 	& 0.81   & 0.23  & 1,2,3\\
NGC~1569  & H {\sc ii}		& IBm 			& 3.63  & 3.36 	& 16.3  & 1.24		& 8.16  & 2.11  &0.0262 & 9.14				& 0.23   & 0.037  & 4,5,6,7,8,9\\ 
NGC~1705  & H {\sc ii}		& SA0$^{-}$ pec 	& 1.91  & 5.10 	& 24.7  & 0.0617*	& 8.48  &0.953 & 1.06    & 8.37$^{\dagger}$	& 0.053 & 0.0084  & 10,11,12,13\\
NGC~1800  & H {\sc ii}		& IB(s)m			& 2.00  & 7.40	& 35.9  & 0.137*	& 8.58  & 1.58  &0.670   & 8.84				& 0.034 & 0.022  & 8,14,15,16\\
NGC~3077  & H {\sc ii}		& I0 pec			& 5.37  & 3.83	& 18.6  & 0.608		& 8.78  & 8.95  &0.0541 & 9.44				& 0.077 & 0.093  & 8,17,18,19,20\\
NGC~5253  & H {\sc ii}		& pec 			& 5.01  & 3.77	& 18.3  & 1.48		& 8.28  & 1.50  &0.0354 & 9.05$^{\dagger}$	& 0.23   & 0.44  & 13,21,22,23,24\\
\enddata

\tablecomments{Column 1: Galaxy Name. Column 2. Optical/UV spectral
  types are identified as either star-forming H {\sc ii} region-like
  galaxies (H {\sc ii}) or contaminated by an active galactic nucleus
  (H {\sc ii} + AGN). Column 3. de Vaucouleurs morphological type from
  \cite{dev91} (hereafter RC3). Column 4. Diameter (major axis) in
  arcminutes based on 25th magnitude B-band observations (RC3). Column
  5. Redshift-independent distance. References: He~2-10:
  \citealt{tul88}; NGC~1569: \citealt{gro08}; NGC~1705:
  \citealt{tos01}; NGC~1800: \citealt{tul88}; NGC~3077:
  \citealt{dal09}; NGC~5253: \citealt{sak04}. Column 6. Spatial scale
  assuming the redshift-independent distance listed in column
  (5). Column 7. Infrared luminosity (8 - 1000 $\mu$m) expressed in
  units of 10$^9$$L_{\odot}$, determined by fitting a three component
  modified blackbody to {\em Herschel} fluxes densities (this work),
  plus {\em Spitzer} and {\em IRAS} flux densities listed in the
  NASA/IPAC Extragalactic Database \citep{mos90,san03}, and the
  distances listed in column (5). The $^*$ indicates that the
  determination of these $L_{IR}$ values included some {\em IRAS} flux
  upper limits. Column 8. Metallicity derived using the method in
  \cite{pet04} (O3N2), which has then been converted via the
  prescriptions in \cite{kew08} to conform with the metallicities in
  \cite{tre04}. Uncertainties in the metallicity come from a
  combination of the method used ($\sim$0.1 dex) and the conversion
  ($\sim$0.06 dex). A solar metalicity [12 + log(O/H)] = 8.69 is
  assumed \citep{all01}. Emission line strength references: He~2-10,
  NGC~1569, NGC~5253: \citealt{kob99}; NGC~1705: \citealt{mou10};
  NGC~1800: \citealt{mou06}; NGC~3077: \citealt{mcq95}. Column
  9. Neutral gas (H {\sc i}) masses based on values from the HIPASS
  \citep{kor04} and THINGS \citep{wal08} surveys and the distances
  listed in column (5). Column 10. Molecular (H$_2$) gas masses
  calculated from CO data from \cite{you95} (He~2-10, NGC~1569,
  NGC~3077, NGC~5253) using the method in \cite{bol13} (eqs. 3 \& 7)
  or by multiplying the star formation rate by an assumed depletion
  time of 2 Gyr (NGC~1705 \& NGC~1800). Column 11. Stellar masses
  calculated using the $M$/$L$ derived from \cite{bel03}, Two Micron
  All Sky Survey $K$ magnitudes \citep{skr03}, and optical colors
  (either $B$-$V$ or $B$-$R$). The ${\dagger}$ indicates masses
  adopted from \cite{zas13} calculated using this method. We used the
  optical colors listed in RC3 to calculate the masses of NGC~1569,
  NGC~1800, and NGC~3077. Column 12. Star formation rate from
  combining $L_{H\alpha}$ \citep{ken08} and $L_{IR}$ (this work, see
  above) according to the prescription in \cite{ken09}
  (eq. 16). Column 13. Star formation rate surface density using the
  star formation rate listed in column (12) and the ionized gas radius
  ($R_{H\alpha}$) listed in \cite{cal10}. Column 14. Selected
  references to a galactic wind or extraplanar diffuse ionized gas
  (eDIG): (1) \citealt{men99}; (2) \citealt{joh00}; (3)
  \citealt{kob10}; (4) \citealt{wal91}; (5) \citealt{hun93}; (6)
  \citealt{hec95}; (7) \citealt{del96}; (8) \citealt{mar97}; (9)
  \citealt{wes08}; (10) \citealt{meu89}; (11) \citealt{meu92}; (12)
  \citealt{meu98}; (13) \citealt{hec01}; (14) \citealt{mar95}; (15)
  \citealt{hun96}; (16) \citealt{ras04}; (17) \citealt{thr91}; (18)
  \citealt{mar98}; (19) \citealt{ott03}; (20) \citealt{cal04}; (21)
  \citealt{cal99}; (22) \citealt{str99}; (23) \citealt{kob08}; (24)
  \citealt{zas11}.}

\end{deluxetable}

\begin{deluxetable}{l c c l l}
\tablecaption{{\em Herschel Space Observatory} Data \label{tbl:data}}
\tabletypesize{\scriptsize}
\tablecolumns{5}
\tablewidth{0pt}
\tablehead{
	\colhead{Galaxy} & 
	\colhead{Instrument} & 
	\colhead{$t_{int}^a$} &
	\colhead{OD/Obs. ID(s)$^b$} & 
	\colhead{Principal Investigator} \\
	\colhead{} &
	\colhead{(PACS$^c$/SPIRE)} &
	\colhead{(hrs)} &
	\colhead{} &
	\colhead{(OT/KPGT/SDP/KPOT)$^{(e)}$}
}
\startdata
He~2-10		& PACS	& 6.56	& 1076/1342244883-89		& Veilleux, S. (OT2) \\
{}			& SPIRE	& 0.07	& 0374/1342196888			& Madden, S. (KPGT) \\
NGC~1569 	& PACS	& 6.56 	& 1058/1342243816-22		& Veilleux, S. (OT2) \\ 
{}			& SPIRE	& 0.15	& 0318/1342193013			& Madden, S. (KPGT) \\
NGC~1705 	& PACS	& 6.56 	& 0968/1342236656-62		& Veilleux, S. (OT1) \\
{}			& SPIRE	& 0.20	& 0158/1342186114			& Madden, S. (SDP) \\
NGC~1800	& PACS	& 6.56	& 1377/1342263895-901$^d$	& Veilleux, S. (OT2) \\
{}			& SPIRE	& 0.16	& 1024/1342240035			& Veilleux, S. (OT2) \\
NGC~3077	& PACS 	& 8.74	& 1059/1342243845-51		& Veilleux, S. (OT2) \\
{}			& SPIRE	& 0.58	& 0318/1342193015			& Kennicutt, R. C., Jr. (KPOT) \\
NGC~5253 	& PACS	& 8.74	& 1194/1342249927-33		& Veilleux, S. (OT2) \\
{}			& SPIRE	& 0.29	& 0459/1342203078			& Madden, S. (KPGT)\\
\enddata
\tablenotetext{a}{Total integration time in hours of the observation.}
\tablenotetext{b}{{\em Herschel} Operation Day / Observation ID number(s).}
\tablenotetext{c}{The PACS photometer scan maps were either 3' scan legs $\times$ 30 legs $\times$ 4" leg separation (He~2-10, NGC~1569, NGC~1705, NGC~1800) or 3' scan legs $\times$ 40 legs $\times$ 4" leg separation (NGC~3077 and NGC~5253).}
\tablenotetext{d}{The PACS photometer observations from OD1377 (NGC~1800) suffered
from a detector problem: After OD 1375 half of the red PACS photometer array was lost. Point-source photometry was still possible, but with lower coverage. Given the small size of NGC~1800, this problem had no impact on the quality of the data.}
\tablenotetext{e}{{\em Herschel} Open Time (OT), Key Programme Guaranteed Time (KPGT), Science Demonstration Phase (SDP), or Key Programme Open Time (KPOT).}
\end{deluxetable}

\begin{deluxetable}{l c c l l l}
\tablecaption{Ancillary Data \label{tbl:ancdata}}
\tabletypesize{\scriptsize}
\rotate
\tablecolumns{6}
\tablewidth{0pt}
\tablehead{
	\colhead{Galaxy} &
	\colhead{Band} & 
	\colhead{$t_{int}^a$} &
	\colhead{Observatory/Instrument} &
	\colhead{Obs. ID(s)$^b$} & 
	\colhead{PI(s)$^c$} \\
	\colhead{} &
	\colhead{} &
	\colhead{(s)} &
	\colhead{} &
	\colhead{} &
	\colhead{}
}
\startdata
He~2-10		& H$\alpha$	& 1200	& Magellan/MMTF		& \nodata		& Oey, S. \\
{}			& 4.5 $\mu$m	& 41.6	& {\em Spitzer}/IRAC	& 4329472	& Rieke, G. \\
{}			& 8.0 $\mu$m	& 41.6	& {\em Spitzer}/IRAC	& 4329472	& Rieke, G. \\
{}			& 24 $\mu$m	& 2.62	& {\em Spitzer}/MIPS	& 4347904	& Rieke, G. \\
\hline
NGC~1569	& H$\alpha$	& 300	& KPNO/Bok			& \nodata 		& Martin, C. L.\\
{}			& 4.5 $\mu$m	& 52		& {\em Spitzer}/IRAC	& 4434944	& Fazio, G. \\
{}			& 8.0 $\mu$m	& 52		& {\em Spitzer}/IRAC	& 4434944	& Fazio, G. \\
{}			& 24 $\mu$m	& 2.62	& {\em Spitzer}/MIPS	& 4435456	& Fazio, G. \\
\hline
NGC~1705	& H$\alpha$	& 1200	& Magellan/MMTF		& \nodata		& Oey, S.\\ 
{}			& 4.5 $\mu$m	& 214.4	& {\em Spitzer}/IRAC	& 5535744, 5536000	& Kennicutt, R. C., Jr. \\
{}			& 8.0 $\mu$m	& 214.4	& {\em Spitzer}/IRAC	& 5535744, 5536000	& Kennicutt, R. C., Jr. \\
{}			& 24 $\mu$m	& 3.67	& {\em Spitzer}/MIPS	& 5549312	& Kennicutt, R. C., Jr. \\
\hline
NGC~1800	& H$\alpha$	& 901.2	& KPNO/Bok			& \nodata		& Martin, C. L.\\
{}			& 4.5 $\mu$m	& 428.8	& {\em Spitzer}/IRAC	& 22530304, 22530560, 22530816, 22531072	& Kennicutt, R. C., Jr. \\
{}			& 8.0 $\mu$m	& 428.8	& {\em Spitzer}/IRAC	& 22530304, 22530560, 22530816, 22531072	& Kennicutt, R. C., Jr. \\
{}			& 24 $\mu$m	& 3.67	& {\em Spitzer}/MIPS	& 22624000	& Kennicutt, R. C., Jr. \\
\hline
NGC~3077	& H$\alpha$	& 1000	& KPNO/Bok			& \nodata		& Kennicutt, R. C., Jr. \\
{}			& 4.5 $\mu$m	& 4758.4 (804)$^d$	& {\em Spitzer}/IRAC	& 4331520, 22000640, 22354944, 22357760,  	& Kennicutt, R. C., Jr., Neff, S., Rieke, G. \\
{}			& {}			& {}		& {}				& 22358016, 22539520, 22539776 & {} \\
{}			& 8.0 $\mu$m	& 4758.4 (804)$^d$	& {\em Spitzer}/IRAC	& 4331520, 22000640, 22354944, 22357760, 		& Kennicutt, R. C., Jr., Neff, S., Rieke, G. \\
{}			& {}			& {}		& {}				& 22358016, 22539520, 22539776 & {} \\
{}			& 24 $\mu$m	& 3.67	& {\em Spitzer}/MIPS	& 17597696	& Rieke, G. \\
\hline
NGC~5253	& H$\alpha$	& 1200	& Magellan/MMTF		& \nodata		& Oey, S., Veilleux, S., Zastrow, J.\\
{}			& 4.5 $\mu$m	& 249.6	& {\em Spitzer}/IRAC	& 4386048	& Houck, J. R. \\
{}			& 8.0 $\mu$m	& 249.6	& {\em Spitzer}/IRAC	& 4386048	& Houck, J. R. \\
{}			& 24 $\mu$m	& 3.67	& {\em Spitzer}/MIPS	& 22679040	& Kennicutt, R. C., Jr.\\
\enddata
\tablenotetext{a}{Total integration time in seconds.}
\tablenotetext{b}{Observation ID number(s).}
\tablenotetext{c}{Principal Investigator(s).}
\tablenotetext{d}{Total IRAC mosaic integration time for NGC~3077 (typical mosaic pixel integration time).}
\end{deluxetable}

\begin{deluxetable}{l c rrrrr c rrrrr c rrrrr}
\tablecaption{Global, Disk, and Circumgalactic Infrared Fluxes$^a$ \label{tbl:flux}}
\tabletypesize{\scriptsize}
\rotate
\tablecolumns{19}
\tablewidth{0pt}
\tablehead{
	\colhead{} & \colhead{} &
	\multicolumn{5}{c}{Global$^b$ (Jy)} &
	\colhead{} &
	\multicolumn{5}{c}{Disk$^c$ (Jy)} &
	\colhead{} &
	\multicolumn{5}{c}{Circumgalactic$^d$ (Jy)} \\
	\cline{3-7} \cline{9-13} \cline{15-19}\\
	\colhead{Galaxy} & \colhead{} &
	\colhead{70} & \colhead{160} & \colhead{250} & \colhead{350} & \colhead{500} & \colhead{} &
	\colhead{70} & \colhead{160} & \colhead{250} & \colhead{350} & \colhead{500} & \colhead{} &
	\colhead{70} & \colhead{160} & \colhead{250} & \colhead{350} & \colhead{500}
}
\startdata
He~2-10		&& 24.206 & 19.260 & 6.696 & 2.543 & 0.878 		&& 24.067 & 18.225 & 6.097 & 2.198 & 0.801 		&& 0.139 & 1.035 & 0.599 & 0.345 & 0.077 \\
NGC~1569 	&& 55.466 & 37.953 & 13.462 & 5.659 & 2.040 	&& 52.945 & 35.007 & 8.399 & 3.043 & 1.183		&& 2.521 & 2.946 & 5.063 & 2.616 & 0.857 \\
NGC~1705 	&& 1.234 & 1.372 & 0.593 & 0.292 & 0.125 		&& 1.082 & 1.142 & 0.458 & 0.227 & 0.107		&& 0.152 & 0.230 & 0.135 & 0.065 & 0.018 \\
NGC~1800	&& 1.048 & 1.900 & 0.909 & 0.436 & 0.182 		&& 0.968 & 1.660 & 0.768 & 0.384 & 0.176		&& 0.080 & 0.240 & 0.141 & 0.052 & 0.006 \\
NGC~3077	&& 17.897 & 22.071 & 10.115 & 4.204 & 1.539 	&& 17.888 & 21.299 & 8.967 & 3.581 & 1.324		&& 0.009 & 0.772 & 1.148 & 0.623 & 0.215 \\
NGC~5253 	&& 30.740 & 21.540 & 8.212 & 3.709 & 1.466 		&& 30.067 & 20.519 & 7.478 & 3.369 & 1.341		&& 0.673 & 1.021 & 0.734 & 0.340 & 0.125 \\
\enddata
\tablenotetext{a}{The tabulated fluxes have a calibration uncertainty of $\pm$10\%.}
\tablenotetext{b}{The global fluxes measured in the 250, 350, and 500 $\mu$m SPIRE maps include background corrections (see \S~\ref{flux_measure}).}
\tablenotetext{c}{The cleaned disk fluxes are calculated as: global fluxes $-$ cleaned circumgalactic fluxes. (see \S~\ref{flux_measure}).}
\tablenotetext{d}{The circumgalactic fluxes are measured from the CLEAN residual maps (see \S~\ref{flux_measure}).}
\end{deluxetable}

\begin{deluxetable}{l c crr c crr c c cc}
\tablecaption{Fits to the Spectral Energy Distributions \label{tbl:sed}}
\tabletypesize{\scriptsize}
\rotate
\tablecolumns{15}
\tablewidth{0pt}
\tablehead{
	\colhead{} & \colhead{} &
	\multicolumn{3}{c}{Global} &
	\colhead{} &
	\multicolumn{3}{c}{Disk} &
	\colhead{} &	
	\multicolumn{1}{c}{Circumgalactic$^a$} &
	\colhead{} &
	\multicolumn{1}{c}{Circumgalactic / Global$^b$}\\
	\cline{3-5} \cline{7-9} \cline{11-11} \cline{13-13}\\
	\colhead{Galaxy} & \colhead{} &
	\colhead{log$\displaystyle{ \left( {M_{global} \over M_{\odot}} \right) }$} & \colhead{$\beta^c$} & \colhead{$T_{dust}$} & \colhead{} &
	\colhead{log$\displaystyle{ \left( {M_{disk} \over M_{\odot}} \right) }$} & \colhead{$\beta^c$} & \colhead{$T_{dust}$} & \colhead{} &
	\colhead{log$\displaystyle{ \left( {M_{cg} \over M_{\odot}} \right) }$} & \colhead{} &
	\colhead{$\displaystyle{ {M_{cg} \over M_{global}} }$} \vspace{2mm}
}
\startdata
He~2-10		&& 6.43 $^{+0.042}_{-0.047}$ & 2.00 & 27.7 $\pm$ 1.2	&& 6.35 $^{+0.062}_{-0.073}$ & 2.00 & 28.6 $\pm$ 1.8	 && 5.64 $^{+100\%}_{-100\%}$ && 0.161 $\pm$ 0.030\\\\
NGC~1569 	&& 5.85 $^{+0.061}_{-0.071}$ & 2.00 & 25.7 $\pm$ 1.4 	&& 5.41 $^{+0.116}_{-0.158}$ & 2.00 & 33.0 $\pm$ 4.5	 && 5.65 $\pm$ 30\% && 0.633 $\pm$ 0.216\\\\
NGC~1705 	&& 5.12 $^{+0.149}_{-0.228}$ & 2.00 & 20.8 $\pm$ 4.2 	&& 5.01 $^{+0.242}_{-0.593}$ & 2.00 & 21.0 $\pm$ 6.7	 && 4.47 $^{+320\%}_{-100\%}$ && 0.222 $\pm$ 0.189 \\\\
NGC~1800	&& 5.65 $^{+0.074}_{-0.090}$ & 2.00 & 20.3 $\pm$ 1.3 	&& 5.62 $^{+0.227}_{-0.504}$ & 2.00 & 19.8 $\pm$ 6.9	 && 4.39 $^{+1200\%}_{-100\%}$ && 0.056 $\pm$ 0.040\\\\
NGC~3077	&& 5.94 $^{+0.045}_{-0.050}$ & 2.00 & 22.8 $\pm$ 1.2 	&& 5.82 $^{+0.040}_{-0.044}$ & 2.00 & 24.0 $\pm$ 0.8	 && 5.29 $\pm$ 58\% && 0.227 $\pm$ 0.033 \\\\
NGC~5253 	&& 5.85 $^{+0.104}_{-0.136}$ & 2.00 & 23.4 $\pm$ 2.2 	&& 5.79 $^{+0.102}_{-0.134}$ & 2.00 & 23.9 $\pm$ 2.1	 && 4.98 $^{+270\%}_{-100\%}$ && 0.133 $\pm$ 0.050 \\
\enddata
\tablenotetext{a}{The circumgalactic dust mass was determined by subtracting the disk dust mass from the global dust mass (see \S~\ref{sed_fit}). The uncertainties on the circumgalactic dust mass come from the uncertainties on the global and disk dust masses added in quadrature.}
\tablenotetext{b}{The uncertainties on the circumgalactic-to-global dust mass ratios come from the {\em fractional} uncertainties on the global and disk dust masses added in quadrature.}
\tablenotetext{c}{The $\beta$ values were fixed at 2.00 (see \S~\ref{sed_fit}).}
\end{deluxetable}

\end{document}